\documentclass{aa}
\usepackage{graphicx}
\usepackage{xcolor}
\usepackage{array}
\usepackage{CJK}
\newcolumntype{P}[1]{>{\centering\arraybackslash}p{#1}}
\usepackage{txfonts}
\newcommand{\erosita}{{eROSITA}\xspace}
\newcommand{\srg}{ {\em SRG}\xspace}
\newcommand{\xmm}{ {\em XMM-Newton}\xspace}

\newcommand{\chandra}{ {\em Chandra}\xspace}
\newcommand{\gaia}{ {\em Gaia}\xspace}
\newcommand{\HB}[1]{#1} 
\newcommand{\TL}[1]{#1} 
\newcommand{\GL}[1]{#1} 
\newcommand{\JS}[1]{#1} 
\newcommand{\KD}[1]{#1} 
\newcommand{\SF}[1]{#1} 
\newcommand{\AGE}[1]{#1} 
\newcommand{\LAST}[1]{#1}
\newcommand{\FINAL}[1]{#1} 
\usepackage{amstext}

\begin{document} 
\begin{CJK*}{UTF8}{gkai}
   \title{The eROSITA Final Equatorial Depth Survey (eFEDS):}
   
   \subtitle{ X-ray catalogue}

    \titlerunning{The eFEDS X-ray catalogue} 
    

   \author{H. Brunner\inst{1}\thanks{email: hbrunner@mpe.mpg.de} \and
            T. Liu\inst{1} (刘腾)\thanks{email: liu@mpe.mpg.de} \and
            G. Lamer\inst{2} \and
            A. Georgakakis\inst{3} \and
            A. Merloni\inst{1} \and
            M. Brusa\inst{4,5} \and
            E. Bulbul\inst{1} \and
            K. Dennerl\inst{1} \and
            S. Friedrich\inst{1} \and
            A. Liu\inst{1} (刘昂)\and
            C. Maitra\inst{1} \and
            K. Nandra\inst{1} \and
            M. E. Ramos-Ceja\inst{1} \and
            J. S. Sanders\inst{1} \and
            I. M. Stewart\inst{1} \and
            T. Boller\inst{1} \and
            J. Buchner\inst{1} \and
            N. Clerc\inst{6} \and
            J. Comparat\inst{1} \and
            T. Dwelly\inst{1} \and
            D. Eckert\inst{7,1} \and
            A. Finoguenov\inst{8} \and
            M. Freyberg\inst{1} \and
            V. Ghirardini\inst{1} \and
            A. Gueguen\inst{1} \and
            F. Haberl\inst{1} \and
            I. Kreykenbohm\inst{9} \and
            M. Krumpe\inst{2} \and
            S. Osterhage\inst{1} \and
            F. Pacaud\inst{10} \and
            P. Predehl\inst{1} \and
            T. H. Reiprich\inst{10} \and
            J. Robrade\inst{11} \and
            M. Salvato\inst{1} \and
            A. Santangelo\inst{12} \and 
            T. Schrabback\inst{10} \and
            A. Schwope\inst{2} \and
            J. Wilms\inst{9}}

   \institute{Max-Planck-Institut f\"ur extraterrestrische Physik, Gie{\ss}enbachstra{\ss}e 1, 85748 Garching, Germany
        \and
            Leibniz-Institut f\"ur Astrophysik Potsdam (AIP), An der Sternwarte 16, 14482 Potsdam, Germany
        \and
            Institute for Astronomy and Astrophysics, National Observatory of Athens, V. Paulou and I. Metaxa, 11532, Greece
        \and
            Dipartimento di Fisica e Astronomia "Augusto Righi", Alma Mater Studiorum Università di Bologna, via Gobetti 93/2, 40129 Bologna, Italy
        \and
            INAF-Osservatorio di Astrofisica e Scienza dello Spazio di Bologna, via Gobetti 93/3, 40129 Bologna, Italy
        \and
            IRAP, Universit\'e de Toulouse, CNRS, UPS, CNES, Toulouse, France
        \and
            Department of Astronomy, University of Geneva, Ch. d'Ecogia 16, 1290 Versoix, Switzerland
        \and
            Department of Physics, University of Helsinki, Gustaf Hällstr\"omin katu 2a, FI-00014 Helsinki, Finland
        \and
            Dr. Karl Remeis-Sternwarte and Erlangen Centre for Astroparticle Physics, Friedrich-Alexander Universit\"at Erlangen-N\"urnberg, Sternwartstra{\ss}e 7, 96049 Bamberg, Germany
        \and
            Argelander-Institut f\"ur Astronomie (AIfA), Universit\"at Bonn, Auf dem H\"ugel 71, 53121 Bonn, Germany
        \and
            Universit\"at Hamburg, Hamburger Sternwarte, Gojenbergsweg 112, D-21029 Hamburg, Germany
        \and
            Institut f\"ur Astronomie und Astrophysik, Universität T\"ubingen, Sand 1, D 72076 T\"ubingen, Germany}

   \date{Received May 7, 2021; accepted June 30, 2021}


  








 
  \abstract
   {The \erosita X-ray telescope on board the {\em Spectrum-Roentgen-Gamma} (\srg) observatory combines a large field of view and a large collecting area in the energy range between $\sim$0.2 and $\sim$8.0\,keV. This gives the telescope the capability to perform uniform scanning observations of large sky areas.}
   {\srg/\erosita performed scanning observations of the $\sim$140 square degree {\em \erosita Final Equatorial Depth Survey} field (the eFEDS field) as part of its performance verification phase ahead of the planned four years of all-sky scanning operations. The observing time of eFEDS was chosen to slightly exceed the depth expected in an equatorial field after the completion of the all-sky survey. While verifying the capability of \erosita to perform large-area uniform surveys and serving as a test and training dataset to establish calibration and data analysis procedures, the eFEDS survey also constitutes the largest contiguous soft X-ray survey at this depth to date, supporting a range of early \erosita survey science \TL{investigations}. Here we {\it i)} present a catalogue of detected X-ray sources in the eFEDS field providing information about source positions and extent, as well as fluxes in multiple energy bands, and {\it ii)} document the suite of tools and procedures developed for \erosita data processing and analysis, which were validated and optimised by the eFEDS work.}
   {The data were fed through a standard data processing pipeline, which applies X-ray event calibration and provides a set of standard calibrated data products. A multi-stage source detection procedure, building in part on experience from \xmm, was optimised and calibrated by performing realistic simulations of the \erosita eFEDS observations. Source fluxes were computed in multiple standard energy bands by forced point source fitting and aperture photometry. We cross-matched the \erosita eFEDS source catalogue with previous XMM-ATLAS observations, which confirmed the excellent agreement of the \erosita and XMM-ATLAS source fluxes. Astrometric corrections were performed by cross-matching the \erosita source positions with an optical reference catalogue of quasars.}
   {We present a primary catalogue of 27910 X-ray sources (542 of which are significantly spatially extended) detected in the 0.2--2.3\,keV energy range with detection likelihoods $\ge 6$, corresponding to a (point source) flux limit of $\LAST{6.5}\times 10^{-15}$ erg/cm$^2$/s in the 0.5--2.0\,keV energy band \HB{(80\% completeness)}.  A supplementary catalogue contains 4774 low-significance source candidates with detection likelihoods between 5 and 6. In addition, a hard-band sample of 246 sources detected in the energy range 2.3--5.0\,keV above a detection likelihood of 10 is provided. In an appendix, we finally describe the dedicated data analysis software package, the \erosita calibration database, and the standard calibrated data products.}
   {}

\keywords{catalogues -- surveys -- X-ray: general}
   
\maketitle
\end{CJK*}
\section{Introduction}

\erosita \citep[extended ROentgen Survey with an Imaging Telescope Array;][]{Predehl2021} is the primary instrument on the {\em Spektrum-Roentgen-Gamma (SRG)} orbital observatory \citep{Sunyaev2021}. It was designed and built, over a period of about 12 years, with the goal of realising a sensitive X-ray telescope with a wide field of view and a significantly larger \HB{grasp}\footnote{In X-ray astronomy, \HB{grasp} is a primary survey metric that is the product of the field-of-view average effective area times the field of view of the telescope.} at 1\,keV than that of either \xmm (by about a factor of 4) or \chandra (by about a factor of 60). These are the two most sensitive focusing X-ray telescopes currently in operation. In addition, a mission plan was devised that included a long (4 years), uninterrupted all-sky survey program (the \erosita All-Sky Survey: eRASS; \citealt{Predehl2021}) in order to guarantee a large volume of accessible discovery space. The \srg is operated by the Space Research Institute of the Russian Academy of Sciences (IKI).

The scientific motivation for the \erosita design was the desire to detect and spatially resolve  a large number (about $10^5$) of clusters of galaxies over a wide redshift range (up to at least $z\sim1$) in order to constrain cosmological parameters \citep[see e.g.][]{aem11,Pillepich2018} at the level of a Stage-IV Dark Energy experiment \citep{Albrecht2006} by characterising the growth of structure.

Following the successful launch in July 2019, and before the start of the all-sky survey, a number of performance verification (PV) observations were carried out
during the early phases of the \srg mission, aimed at verifying all different aspects of the instrument capabilities. These early post-commissioning phases of science operations have clearly demonstrated that \erosita is capable of delivering the design performance in terms of sensitivity, image quality, and spectroscopic capabilities \citep[see e.g.][]{Merloni2020, Predehl2021}. Nevertheless, the ambitious scientific goals of \erosita, and in particular its cosmological objectives, require accurate control over a number of elements, including the modelling of the instrumental and astrophysical backgrounds, the reconstruction of the selection function for both point-like and extended X-ray sources over very large sky areas, the identification of reliable counterparts of the X-ray sources at longer wavelengths, the measurement of their distances (redshifts), the calibration of the empirical (or physical) relations between X-ray observables (e.g. cluster luminosities, temperatures, and density profiles) and cosmologically meaningful parameters, particularly cluster masses.   

 The {\em\erosita Final Equatorial Depth Survey} (eFEDS) was the largest investment of observing time during the PV phase (about \HB{360\,ks}, or 100 hours, in total), and was designed to test the key elements of the science workflow from X-ray photon detection to astrophysics and cosmology.
The eFEDS field, an area of approximately 140 deg$^2$ composed of four individual rectangular raster-scan fields of $\sim$35 deg$^2$ each, was chosen because its \FINAL{multi-wavelength} coverage for an extragalactic field of this size is one of the richest, and because it was visible for \erosita in the limited time period allocated to PV observations. The field \TL{coincides} with an area
 enriched by deep \FINAL{optical and near-infrared (NIR)} imaging of the HSC\footnote{\FINAL{Hyper Suprime-Cam; https://subarutelescope.org/}} Wide area Survey \citep{Aihara2018}, KIDS-VIKING\footnote{\FINAL{https://kids.strw.leidenuniv.nl/}} \citep{Kuijken2019}, DESI Legacy Imaging Survey\footnote{\FINAL{https://www.legacysurvey.org/}} \citep{Dey2019}, and, among others, GAMA\footnote{\FINAL{http://www.gama-survey.org/}} \citep{Driver09}, \FINAL{WiggleZ} \citep{Drinkwater2018}, LAMOST\footnote{http://dr5.lamost.org/doc/release-note-v3}, and the SDSS\footnote{\FINAL{Sloan Digital Sky Survey; https://www.sdss.org/}} \citep{Blanton2017} spectroscopic coverage. A full description of the available multiwavelength data, together with the identification of the multiwavelength counterparts, is presented in \cite{Salvato2021}. The field also contains a medium-wide \xmm survey field (the XMM-ATLAS; \citealt{Ranalli2015}), which provides a useful dataset for a comparison and validation of the \erosita X-ray analysis pipeline.

The eFEDS observational strategy was designed so as to provide uniform exposure over the field about 50\% deeper than what is expected for eRASS at the ecliptic equator (i.e. over most of the sky) at the end of the 4-year all-sky survey program, while at the same time covering a sufficiently wide area to provide large statistical samples of different source classes. In particular, enough clusters were to be provided for calibrating the mass-observable relation using the weak-lensing maps provided by the exquisite optical imaging of the HSC survey. 

In a series of accompanying papers, we will present the identification of the multiwavelength counterparts of point-like \LAST{\citep{Salvato2021,Schneider2021}} and extended \LAST{\citep{Klein2021}} X-ray sources; the resulting clean catalogues of clusters of galaxies \LAST{\citep{Liu2021_cluster,Bulbul2021}}, active galactic nuclei \FINAL{(AGN)} \LAST{\cite[][Nandra et al., in prep.]{Liu2021_AGN}} and stars \LAST{\citep{Schneider2021}}; the X-ray variability properties of the detected sources \LAST{\citep{Boller2021,Buchner2021}}; the X-ray spectral analysis of the point sources \citep{Liu2021_AGN}; the X-ray morphological analysis of the clusters \LAST{\citep{Ghirardini2021b}}; the optical and lensing analysis of the X-ray clusters \LAST{\citep[][Ota et al., in prep.]{RamosCeja2021,Chiu2021,Bahar2021}}; the properties of non-active galaxies detected by \erosita \LAST{\citep{Vulic2021}}, and the discovery of some extreme AGN \LAST{\citep{Toba2021,Brusa2021}}. Earlier works based on the eFEDS data have been published and include the discovery of a supercluster \citep{Ghirardini2021} and a very high-redshift quasar \citep{Wolf2021}.

In this paper, we describe in detail the X-ray observations, the data analysis and calibration procedures, the algorithms we used to detect and characterise X-ray sources, and  their validation through an extensive simulation analysis. We also release here the resulting catalogues of X-ray sources, and present the description of the dedicated software system (the \erosita Science Analysis Software System; eSASS) in a series of appendices, which is also made public concurrently with this paper. The appendices include a basic description of the operating principles and algorithms of the data analysis pipeline. A full description of the 
eSASS software is also available online\footnote{\HB{https://erosita.mpe.mpg.de/edr/DataAnalysis/\label{webdoc}}}.

\section{Observations}

\begin{figure*}
    \centering
 \includegraphics[width=0.9\textwidth]{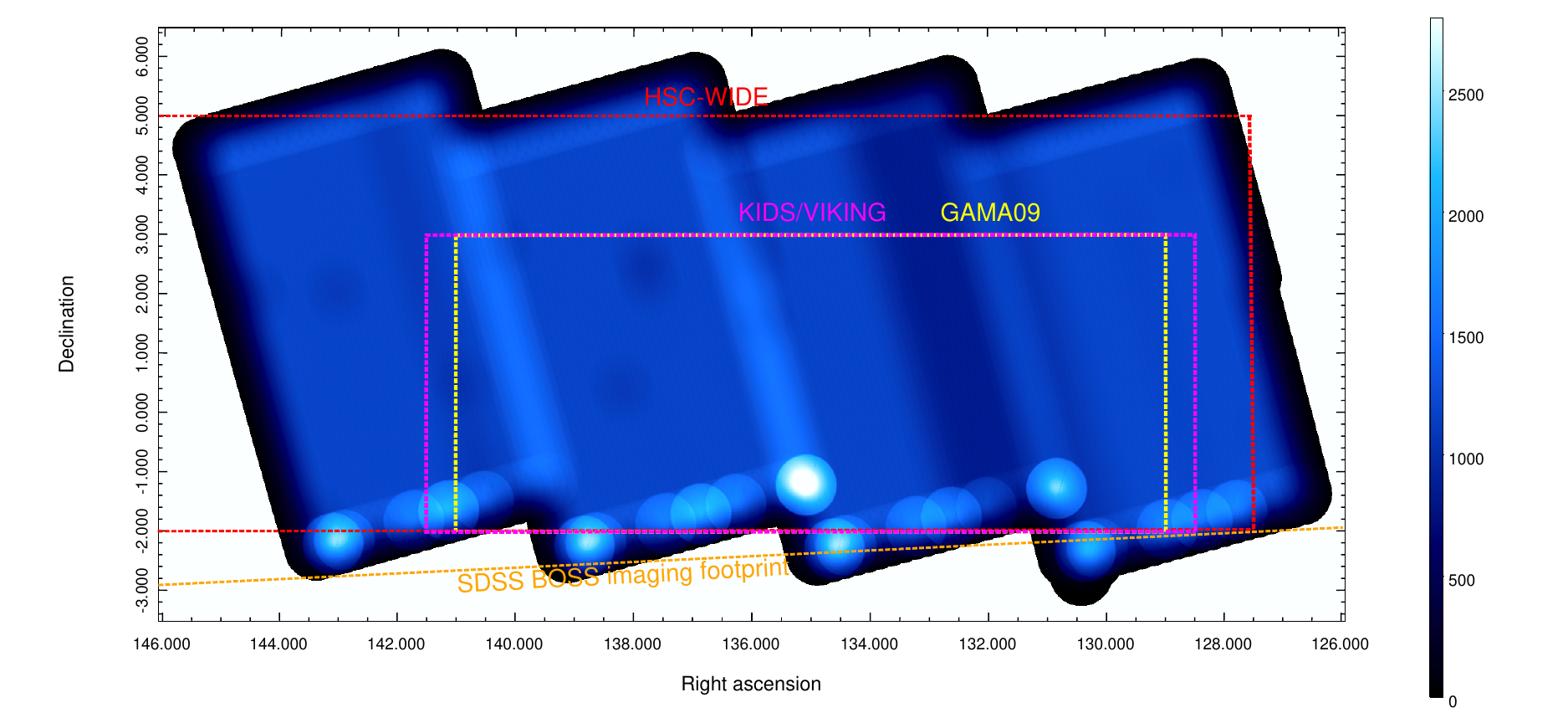}
 \includegraphics[width=0.82\textwidth]{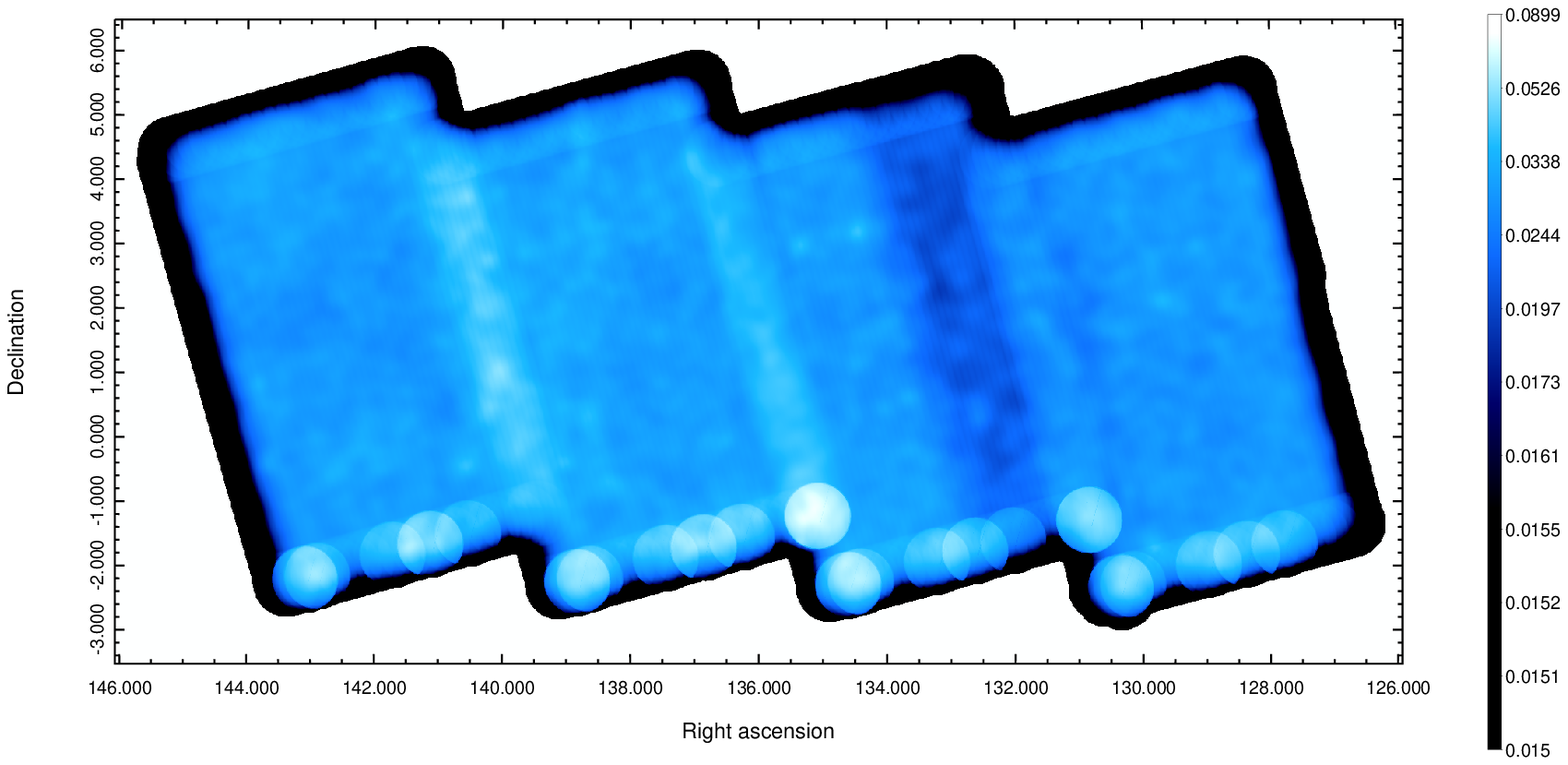}
    \caption{ 0.2-2.3\,keV exposure map (vignetting corrected, in units of seconds; upper panel) and background map (in units of counts per pixel; lower panel). The bright circles mark high-exposure regions and are a result of the scanning mode used in the observations. They are created due to the waiting time of the spacecraft before inverting its scanning direction. The darker stripe in the second rectangular chunk on the right is due to the malfunction of two of the seven \erosita cameras during that period. The upper panel also shows the footprints of a few relevant \FINAL{optical and NIR, imaging and spectroscopic} surveys. 
    }
    \label{fig:exp_bkg}
\end{figure*}

The \srg mission supports three principal observing modes \citep{Predehl2021,Sunyaev2021}. As its main objective, \srg is in the process of performing a 4-year survey of the full sky in continuous scanning mode, in which the spacecraft pointing direction traces great circles in the sky with a rotation speed of 90 degrees per hour (or $\sim 90\arcsec/$s). In addition, \srg is capable of making pointed observations of individual targets, as well as extended rectangular fields in so-called field-scanning mode. The pattern of the field-scanning mode consists of parallel scans in alternating directions, offset by \HB{$6\arcmin$}. Given the \erosita field of view ($\approx 1$ degree in diameter), each sky position is thus observed in ten consecutive scans. The maximum supported size of the scanned rectangles is $12.5^\circ \times 12.5^\circ$. In the interest of operational simplicity, the orientation of the rectangles is aligned with the ecliptic coordinate system. For mission planning purposes, individual field-scan observations are scheduled to be not significantly longer than one day, thus fitting between consecutive ground contacts with the SRG spacecraft. These constraints necessitate splitting the observations of the eFEDS field into four sub-fields of size $4.2^\circ \times 7.0^\circ$ (this is the area with the nominal exposure depth; the area is larger by $\sim0.5^\circ$ on each side, but the exposure is reduced). A scanning speed of $13\farcs15/$s results in a uniform exposure depth of $\sim2.2$~ks ($\sim1.2$~ks after correcting for telescope vignetting) across most of the field. 
Each source is observed continuously for up to 4.7 min in an individual scan when passing through the central part of the field of view and for shorter time periods in peripheral scans. 

The eFEDS observations were carried out at the beginning of November 2019 with all seven \erosita Telescope Modules (TM1-7) in operation. However, due to an unrecognised malfunction of the camera electronics, 28\% of the TM6 data of eFEDS sub-field I and 48\% and 43\% of the TM5 and TM6 data, respectively, of eFEDS sub-field II could not be used. This resulted in a reduced exposure depth in the affected areas of up to $\sim$30\% (as visible in Figure~\ref{fig:exp_bkg}).

We report an unrecognized calibration error \LAST{of 1\,s} in the arrival time correction \LAST{for} approximately half of the TM6 data of eFEDS sub-field I prior to the malfunction. Because of the scanning pattern of the eROSITA field-scan observing mode, this causes photons observed by this camera to be projected onto the sky with an offset of $\sim$13\arcsec on either side of the true position, resulting in an elongation of the merged all-camera \FINAL{point spread function (PSF)} in the scanning direction in the affected area on the eastern side of sub-field I. Some technical details about the cause and correction method of these 1\,s time shifts are provided in Appendix B.3.

All \erosita cameras are protected by light-blocking filters. For five of the seven cameras, a 200\,nm Al layer is deposited directly on the CCDs\footnote{\FINAL{charge-coupled device sensors of the \erosita cameras}} , while for the remaining two (TM5 and TM7), the suppression of optical light is achieved by a 100\,nm Al layer on an external filter in the filter wheel. After launch it was found that a small amount of scattered sunlight can reach the CCDs from the side, bypassing the filter wheel \citep['light leak'; see][]{Predehl2021}. This causes a spatially inhomogeneous raised background at low energies in TM5 and TM7 that changes with the orientation of the spacecraft with respect to the Sun. 

\KD{This light-leak noise, which is smoothly distributed at scales $>$10\arcmin, will be included in the background map and thus has a minor impact on the source detection.
The optical photons may also cause an energy shift that is negligible for the source detection, however. Therefore, all seven cameras are used in this work.
}


Table~\ref{tab:obs} presents the details of the eFEDS field observations. The top panel of Figure~\ref{fig:exp_bkg} shows the final exposure map of the eFEDS survey in the soft band (0.2--2.3\,keV), corrected for vignetting\footnote{\TL{As the eFEDS field is scanned uniformly, the ratio of the unvignetted exposure time and the 0.2-2.3 keV vignetted exposure time is almost a constant (1.86) across the field, except at the field border.}}. The schematic footprints of a few relevant \FINAL{optical and NIR, imaging and spectroscopic} surveys are overlaid on this map.

\begin{table*}[htbp]
    \centering
    \caption{\erosita observations of the four sub-fields of eFEDS}
    \begin{tabular}{c|lllllll}
    \hline
  &     ObsID  &        Central R.A. &  Central Dec & t$_\mathrm{exp}$ & Start time [UTC] & $\Delta \alpha$ & $\Delta \delta$ \\
  &            &        [deg]       &  [deg]      & [s]              &                  &   [arcsec]   & [arcsec] \\
  \hline
I &     300007 &        129.55 &        +1.5 &   89642 & 2019-11-03T02:25:50 &   5.0 & -4.2 \\
II &    300008 &        133.86 &        +1.5 &   89642 & 2019-11-04T04:05:52 &   5.0 & -3.2 \\
III &   300009 &        138.14 &        +1.5 &   89642 & 2019-11-05T05:45:54 &   5.1 & -3.5 \\
IV &    300010 &        142.45 &        +1.5 &   89642 & 2019-11-06T07:25:56 &   4.6 & -4.0 \\
\hline
    \end{tabular}
    \tablefoot{$\Delta \alpha$, $\Delta \delta$: the corrections that were applied to the raw attitude solutions for each ObsID in order to remove systematic linear offsets of derived X-ray source positions in right ascension (R.A.; $\alpha$) and declination (DEC; $\delta$), using the \gaia-unWISE AGN candidate catalogue of \citet[][]{Shu2019} as an astrometric reference. See section \ref{preproc} for details.}
    \label{tab:obs}
\end{table*}

\begin{figure*}
    \centering
\includegraphics[trim=250 0 0 0, clip,width=\textwidth]{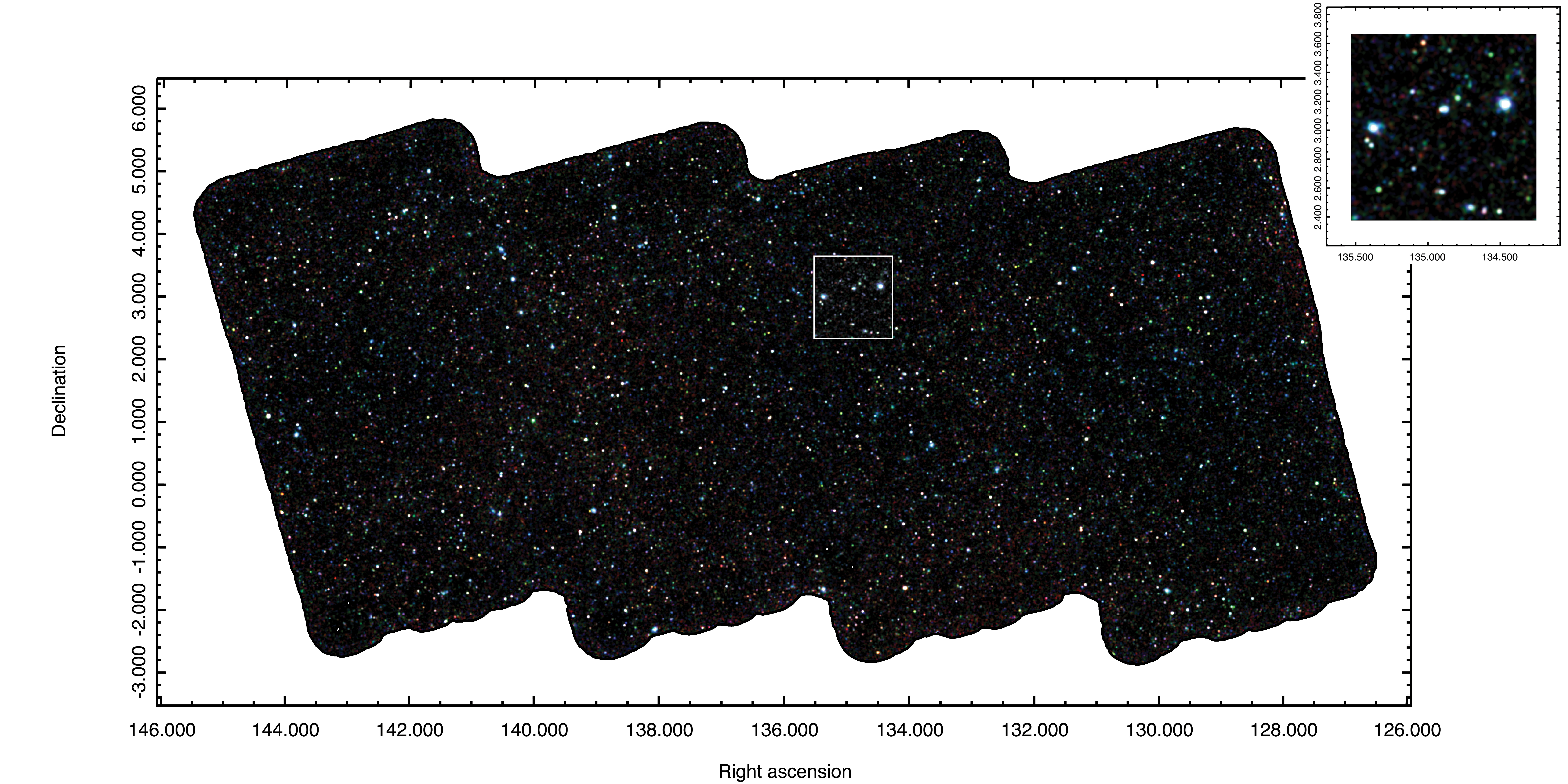}
    \caption{RGB image of the eFEDS field in X-rays, created using 0.2--0.5 (R), 0.5--1 (G), and 1--2 (B)\,keV bands. Each count rate image has been smoothed with a Gaussian with $\sigma=10$~pixels (40\arcsec).
The inset in the upper right corner shows a zoom-in around a newly discovered supercluster \citep{Liu2021_cluster}.}
    \label{fig:image}
\end{figure*}

\TL{Figure~\ref{fig:image} shows} an exposure-corrected RGB image of the entire field. The images were created using the 0.2--0.5\,keV (red), 0.5--1\,keV (green), and 1--2\,keV (blue) energy bands, after smoothing each count rate image with a Gaussian with $\sigma=40\arcsec$.

\section{Data analysis}
In this section we describe the details of the data processing with the \erosita eSASS. The eSASS data analysis package and its tasks are described in detail in Appendix~\ref{appendix_esass}. Further documentation including a description of all command-line parameters is available online\textsuperscript{\ref{webdoc}}.

  \subsection{Data reception and standard pipeline processing}
  \label{preproc}
  \erosita science and auxiliary telemetry data received during each daily SRG ground contact are transferred from the IKI to the \erosita Data Centre at the Max Planck Institute for Extraterrestrial Physics (MPE) in Garching, Germany, where they undergo various processing steps as part of a standard data analysis pipeline. The data are decommutated, \TL{packaged}, and converted into FITS format \citep{Wells1981} files in an initial pre-processing step. They are subsequently \TL{fed} through an event-processing task chain that creates calibrated X-ray event files suitable for scientific analysis. The main software tasks of the event-processing chain are described in Appendix \ref{appendix_esass}.1 and \ref{appendix_esass}.2.; calibrated event files are described in Appendix \ref{appendix_products}.1. In addition, the standard data processing pipeline provides a range of high-level data products such as exposure, background, and sensitivity maps, as well as X-ray source catalogues and source-specific products such as spectra and light curves. They are described in Appendix \ref{appendix_products}. 
  
  This work is based on the pipeline-\HB{generated} calibrated event files, while higher-level data products are created by \TL{an additional} pipeline. The data were processed using pipeline version c001, which was released as part the \erosita Early Data Release\footnote{\HB{available for download at\newline https://erosita.mpe.mpg.de/edr/eROSITAObservations/}}. 
   
   \subsection{Astrometric corrections and data preparation}
   \label{sec:firstastrocorrection}
   
   Some artefacts in the data due to a temporary malfunctioning of the camera electronics of TM4 are not yet perfectly removed in this version of the pipeline. In the TM4 data of all the four observations, we therefore removed two bright pixels (pixel coordinate \texttt{RAWX}, \texttt{RAWY}: 115, 235, and 178, 225). Furthermore, we removed the soft photons with a \texttt{PI} (event energy in eV) below 600 and \texttt{RAWY} above 120 in a few columns (\texttt{RAWX}: 121, 125, 127, 259, 262, 380, and 382).  The fraction of removed events is negligible.
   
We carried out an initial round of astrometric corrections of the X-ray dataset as follows. We created images and detected sources in the 0.2--2.3\,keV band separately for each of the four eFEDS ObsIDs (using the methods described below). For each ObsID, \TL{we searched for the closest \FINAL{optical and IR} AGN in the \gaia-unWISE AGN catalogue \citep{Shu2019} for each detected point \LAST{source} (\texttt{EXT}=0) within a maximum separation of 30\arcsec using an iterative \TL{$3\sigma$ clipping} algorithm. We found 1600--1900 matches per ObsID and} calculated the right ascension (RA; $\alpha$) and declination (DEC; $\delta$) offsets $\Delta \alpha$, $\Delta \delta$ for each ObsID. This removes the mean linear offsets of the X-ray sources from the \gaia DR2 positions. We applied the computed corrections (see Table~\ref{tab:obs}) to the observation attitude and then recalculated the event coordinates using the tasks \texttt{evatt} and \texttt{radec2xy} (see Appendix~\ref{app:att}).

After applying astrometric corrections to each of the four observations separately, we merged them into one \FINAL{and applied} filters of \texttt{FLAG=0xc00fff30} (this selects good events from the nominal field of view, excluding bad pixels) and \texttt{PATTERN$\le 15$} (this includes single, double, triple, and quadruple events). In the merged events, we searched for background flares by running \texttt{flaregti} (see Appendix~\ref{app:photometry}) in the 0.2--5\,keV band\footnote{\TL{For \texttt{flaregti}, we adopt a source diameter of 120\arcsec, a source likelihood threshold of 10, a time bin size of 100s, and 60 grid points per dimension.}}. Only one short ($<1$ ks) significant flare was detected by \texttt{flaregti} in the entire dataset \TL{(algorithm described in Appendix \ref{app:photometry})}. 
We applied the \texttt{flaregti} filter with \texttt{evtool} (see Appendix~\ref{app:events}) to filter the background flare out.

Using \texttt{evtool}, we extracted the events in specific energy ranges and created images with a resolution of $4\farcs0$ per pixel. This is a factor 2.4 higher than the size of the physical pixels of the \erosita cameras ($9\farcs6$).
We created vignetted exposure maps in each band and an unvignetted exposure map using \texttt{expmap}.
A source detection mask was created with \texttt{ermask} on the basis of the 0.2--2.3\,keV band vignetted exposure map by applying a minimum cut at 1\% of the maximum value. 
\TL{By adopting such a low exposure threshold, we included the field border in the analysis. The depth is significantly shallower than the main field in the border. These shallow border regions can be excluded when necessary (\S~\ref{sec:lognlogs}).}

\subsection{Source detection}
\begin{figure}
  \centering
   \includegraphics[width=0.8\columnwidth]{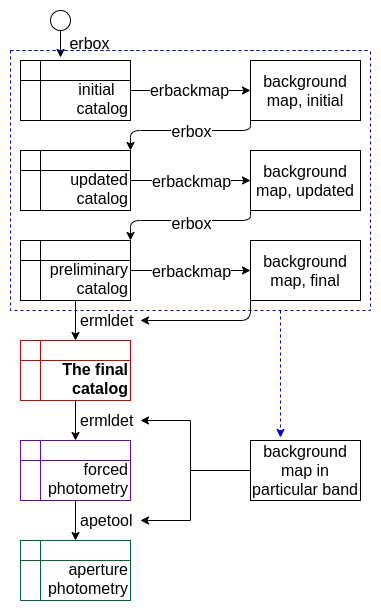}
   \caption{Source detection procedures.}
   \label{fig:flowchart}
\end{figure}

We created the main eFEDS catalogue by  running a single-band source detection in the 0.2--2.3\,keV band using all TMs. This \TL{band} guarantees the highest sensitivity \FINAL{given} the shape of the \erosita response \citep{Predehl2021}. This was shown by simulation tests \citep{Liu2021_sim}.
In addition, we also ran the same source detection procedure simultaneously in three bands (0.2--0.6, 0.6--2.3, and 2.3--5\,keV), in order to select sources with particularly hard (or soft) spectra. 
\TL{As the eSASS tasks are able to handle either one set of files (e.g. event file, image, or background map) in a particular energy band or multiple sets of files in a few bands, the source detection procedure is identical for the single-band detection and the three-band detection.
The source detection procedure is illustrated in Figure~\ref{fig:flowchart} and is described below. We present the adopted values of the key parameters of the eSASS tasks we used here. They are described in more detail in Appendix~\ref{appendix_esass}.
}

 \begin{figure}
     \centering
          \includegraphics[width=\columnwidth]{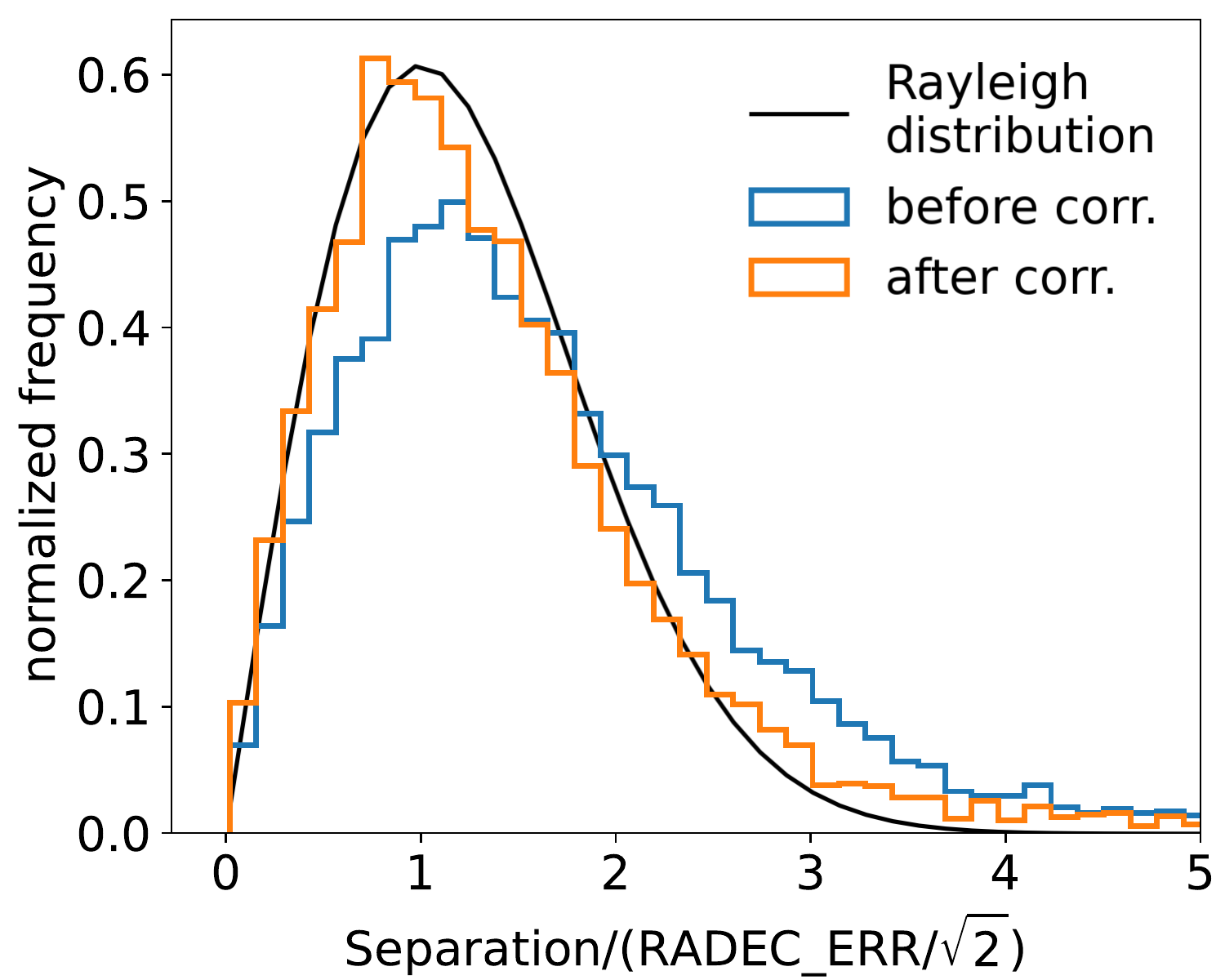}

     \caption{Distributions of the separations between the optical positions of the \gaia-unWISE AGN catalogue and the X-ray positions before and after the second-pass corrections. 
     }
     \label{fig:Rayleigh}
 \end{figure}
 
 \begin{figure}
     \centering
     
     \includegraphics[width=\columnwidth]{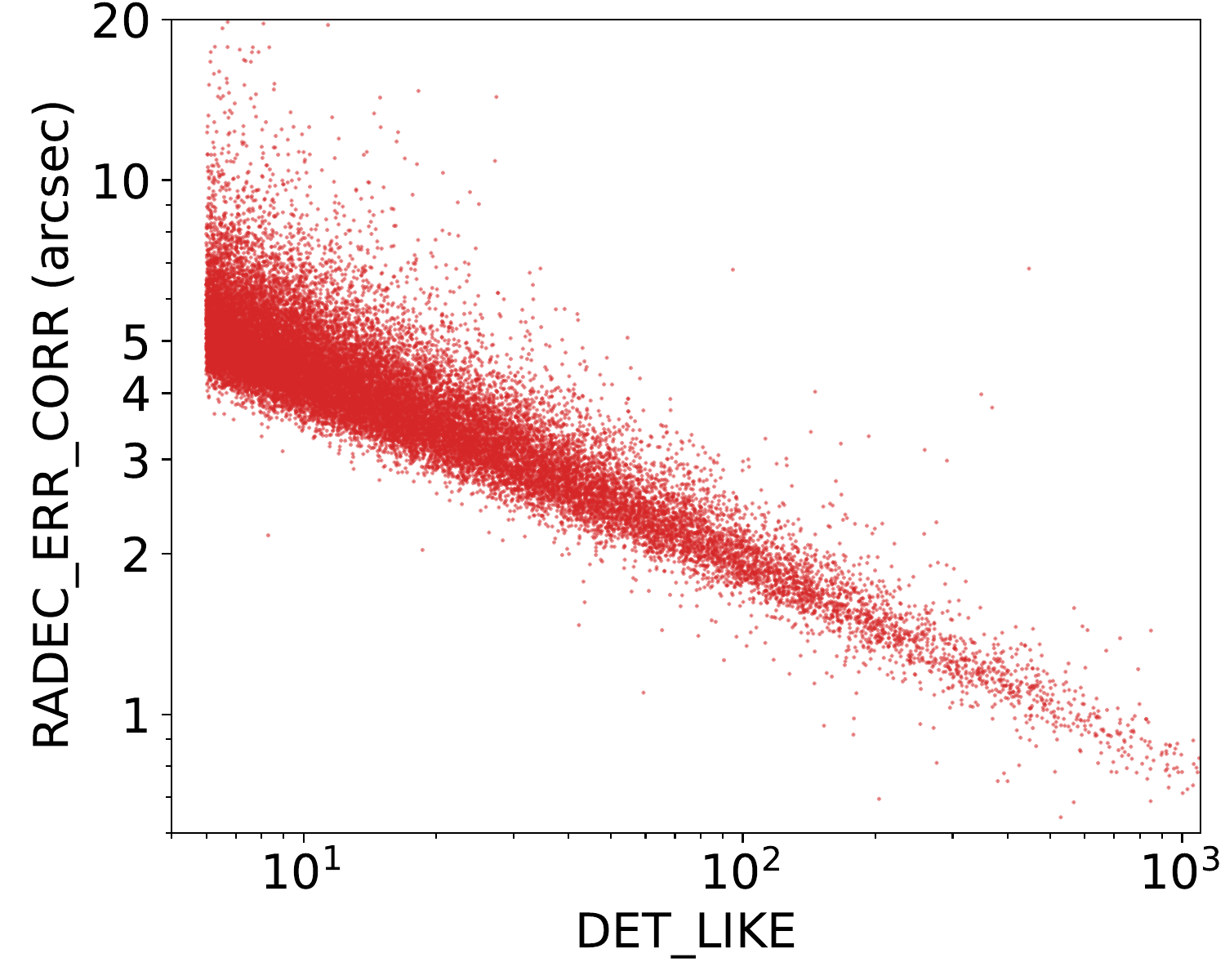}
     \caption{Positional uncertainty (after astrometric correction, \texttt{RADEC\_ERR\_CORR}) of all point sources in the main catalogue as a function of the detection likelihood in the 0.2-2.3\,keV energy band.
     }
     \label{fig:rates_error}
 \end{figure}

\TL{
Our core source detection algorithm selects source candidates according to the statistics of fitting source images with a PSF-convolved model (beta model or $\delta$ function), which is called PSF-fitting hereafter. Before the PSF-fitting, a preliminary catalogue that contains all the potential source candidates must be prepared.
To prepare thispreliminary catalogue, a background map must be available.
We therefore first ran \texttt{erbox}\footnote{\TL{For \texttt{erbox}, we always adopted two-step image rebinning (\texttt{nruns}=2).}} in local mode (without background map) and adopted a detection likelihood threshold of 6 and a box size of 9 pixels (\texttt{boxsize}$=$4) to create an initial catalogue.
This initial catalogue, in which the quality is not controlled, was only used to create a background map using \texttt{erbackmap} (adopting a likelihood threshold of 6 and a required signal-to-noise ratio of 40), which masks out the sources in the catalogue and then adaptively smooths the image to create the background map.
After preparing the background map prepared, we ran \texttt{erbox} in map mode, adopting a detection likelihood of 4 and a box size of \LAST{9} pixels, to create an updated catalogue.
Considering that this updated catalogue is affected by the settings that were adopted when the initial catalogue was adopted through the background map,
we repeated the step of creating a background map and running \texttt{erbox} with the same parameters and updated the catalogue again, creating the preliminary catalogue.
This preliminary catalogue is determined by the parameter settings used in the background map \TL{creation} and the map-mode \texttt{erbox} detection.
Using the preliminary catalogue, we created a final background map using the same parameters as above. The bottom panel of Figure~\ref{fig:exp_bkg} shows the resulting background map in the 0.2-2.3~keV band.
}

Finally, the preliminary catalogue was used as input to the PSF-fitting in the next step, which selects reliable sources from this catalogue.
We ran photon-mode PSF-fitting using \texttt{ermldet} and adopted a PSF-fitting radius (\texttt{cutrad}) of 15 pixels, a multiple-source searching radius (\texttt{multrad}) of \LAST{20} pixels, a detection likelihood threshold (\texttt{likemin}) of 5, an extent likelihood threshold (\texttt{extlikemin}) of 6, an extent range between 2 and 15 pixels, a maximum of {\LAST four} sources for simultaneous fitting, and allowed a source to be split into two sources at most.
We finally used \texttt{catprep} to format the final catalogue.
The preliminary catalogue (containing 84565 sources in the single-band detection and 58227 sources in the three-band detection) is much larger than the PSF-fitting output catalogue.
As mentioned before, the PSF-fitting itself was performed within a circular region with a radius that was fixed by \texttt{cutrad}, which is an essential parameter of \texttt{ermldet}. Through simulation tests, we determined that \texttt{cutrad=15} is a good choice. Adjusting it does not improve the detection efficiency of point sources significantly.

\TL{
By comparing the best-fit source model with a zero-flux (pure background) model, \texttt{ermldet} calculates a detection likelihood (\texttt{DET\_LIKE}) $L$ for each source, defined as $L=-\ln P$, where P is the probability of the source being caused by random background fluctuation.
By comparing the extended beta model with a $\delta$ function, \texttt{ermldet} also calculates an extent likelihood \texttt{EXT\_LIKE} for each source, which is defined corresponding to the probability of a source being unresolved rather than extended. Sources with an extent likelihood above and below \texttt{extlikemin} (6) were fitted with the beta model and $\delta$ function, respectively.} \GL{These sources have catalogue values \texttt{EXT}$=$0.0 and \texttt{EXT\_LIKE}$=$0.0. }  
\TL{A low threshold of \texttt{extlikemin}$=$6 was chosen in order to achieve a high completeness of the extended source sample, allowing some point sources to be misclassified as extended \citep{Liu2021_sim}. 
To increase the completeness of the point source catalogue or the purity of the extended source catalogue, a higher cut on \texttt{EXT\_LIKE}, such as 8 or 12, can be applied when needed.
By minimising the C-statistic, \texttt{ermldet} measures the source position, extent, and count rate, together with the $68\%$ confidence intervals. The task determines error margins by varying each fit parameter from the best-fit position in both directions  to find the points at which the likelihood function $\Delta C=C - C_\mathrm{best}$ reaches the value 1.0. The error values from both directions are averaged and a single error value is written to the output file.
For this reason, the error values and the combined errors for faint sources with large uncertainties have to be treated with caution.
In the case of multi-band detection, the total-band count rates \texttt{ML\_RATE\_0}, counts \texttt{ML\_CTS\_0,} and fluxes \texttt{ML\_FLUX\_0} have the errors of the single-band rates, counts, and fluxes added in quadrature, treating them as Gaussian errors. 
The algorithm of \texttt{ermldet} is described in more \LAST{detail} in Appendix \ref{appendix_esass}.5. }

\TL{For the single-band detection, the 0.2-2.3~keV event file, image, and vignetted exposure map were used.
For the three-band detection, the background maps were created using the same masking catalogue, but separately in the three bands; and the event files, images, exposure maps, and background maps of the three bands are input simultaneously into \texttt{erbox} and \texttt{ermldet} for source detection.
The vignetted exposure map was used except for the 2.3--5\,keV band, where the unvignetted exposure map was used instead because \TL{the background in the hard band is dominated by high-energy particles \citep{Predehl2021}.}}

\TL{
After source detection, we matched the X-ray sources to the \gaia-unWISE AGN catalogue \citep{Shu2019} , for which we again adopted a maximum separation of 30\arcsec . we display the X-ray to optical positional separation in Figure~\ref{fig:Rayleigh}.
We made a second-pass astrometric correction of the catalogue in right ascension and declination, allowing a linear shear term rather than considering only linear shift (as was done in \S~\ref{sec:firstastrocorrection}). Simulations show that the direct measurements of the positional uncertainty are slightly underestimated \citep{Liu2021_sim}.}
We therefore also computed an empirical correction of the raw positional uncertainty estimates $\Delta \theta$ , that brings the distribution of the observed X-ray to reference catalogue offsets closer to the expected Rayleigh distribution. The computed corrections are as follows:
\begin{eqnarray}
\alpha_{\rm CORR} = \alpha-(-0.2158 \times \delta+0.4526)/\cos(\delta)/3600 \nonumber \\
\delta_{\rm CORR} = \delta-(0.086 \times \delta + 0.0679)/3600 \nonumber \\
\Delta\theta_{\rm CORR} = 1.15\times\sqrt{\Delta \theta^2+0.7^2} 
 .\end{eqnarray}

\begin{figure*}
     \centering
     \includegraphics[width=1.8\columnwidth]{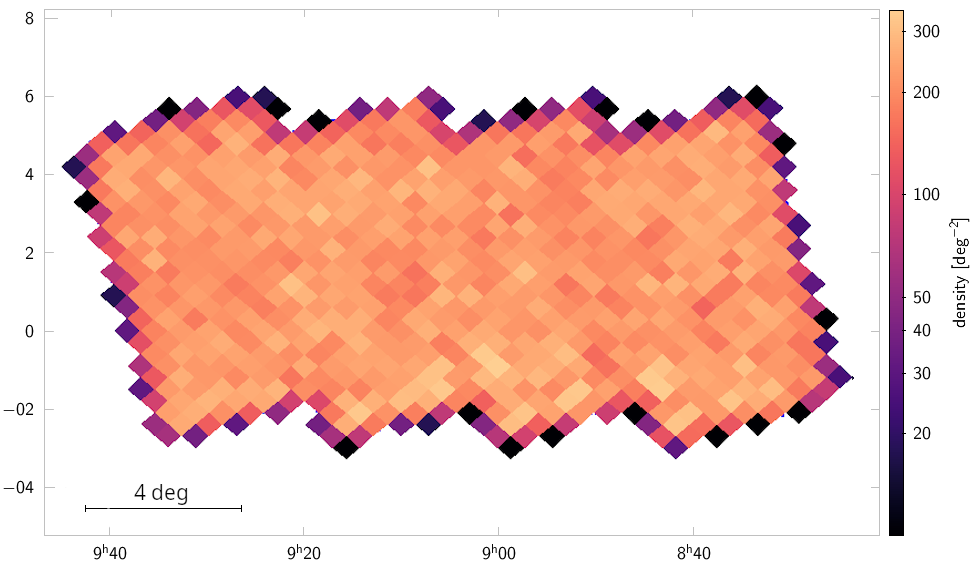}
     \caption{\LAST{HEALpix} map (\texttt{Nside=128}) showing the projected sky density of the eFEDS main catalogue point sources in equatorial coordinates. Each pixel has a size of $\approx$0.2098 deg$^2$.
     }
     \label{fig:density}
 \end{figure*}

 The second-pass astrometric correction results in a much better consistency between the separation distribution and the Rayleigh distribution, although
 \TL{a tail beyond the Rayleigh distribution is still visible at large separations. This is mainly caused by faint sources, which have relatively higher probabilities to be spurious and relatively larger positional uncertainties, and could more easily have underestimated positional uncertainties or false optical counterparts.}
 The mean and median positional uncertainty after this correction are $4\farcs73$ and $4\farcs65$, respectively. 
 Figure~\ref{fig:rates_error} shows the corrected positional uncertainty as a function of the source detection likelihood in the 0.2--2.3\,keV band. $\alpha_{\rm CORR}$, $\delta_{\rm CORR}$, and $\Delta\theta_{\rm CORR}$ are added to the source catalogues as columns \texttt{RA\_CORR}, \texttt{DEC\_CORR}, and \texttt{RADEC\_ERR\_CORR}, respectively.

\subsection{Forced photometry}
\label{sec:forced_phot}
We ran forced PSF-fitting photometry in seven energy bands: 0.2--0.5, 0.5--1, 1--2, 2--4.5, 0.5--2, 2.3--5, and 5--8\,keV (Table~\ref{table:bands}). \TL{The procedure is described as follows.}

\TL{Before the forced PSF-fitting,  a background map has to be created for each band.
We used the same method as described above for the source detection band, as illustrated in the dashed blue box in Figure~\ref{fig:flowchart}.
The only difference is that we did not need to create an initial catalogue again, as we were able to use the preliminary catalogue generated in the single-band detection.
Running the \texttt{erbackmap} background map creating and the \texttt{erbox} detection procedure twice and using the same parameters as above, we created a preliminary catalogue for each band.
Using this catalogue, we created the final background map for each band.
Instead of creating a preliminary catalogue as described above for each band, an even simpler method is using the single-band detected catalogue directly, which will lead to similar results. However, we adopted this more complex method because of the advantage of having the background in each band determined by only the data in the same band. In this way, the background only depends on the adopted task parameters.
}

The preliminary catalogue of each band was only used to create the background map. When this was done, we input the full single-band detected (or three-band detected) catalogue into \texttt{ermldet} \footnote{\TL{We used \texttt{ermldet}-1.47 here because a bug that affects the forced PSF-fitting in the c001 version of \texttt{ermldet} is solved in this updated version.}} PSF-fitting, but fixed the source position and extent.
This forced PSF-fitting only provides a count rate measurement for each of the chosen bands.
The results of the forced fitting are given for all input sources, but we note that many of the single-band measurements are not significant, in particular for the hard bands beyond 2\,keV.

For the two harder bands (2.3--5 and 5--8\,keV), we used an unvignetted exposure map to generate the background map. In the forced PSF-fitting, however, the vignetted exposure map was used, so that the measured count rate is corrected for vignetting. Fluxes in each forced photometry energy band were computed by assuming a power-law spectrum with a spectral index $\Gamma=2.0$ and a galactic absorption of $N_H=3\times 10^{20}$ cm$^{-2}$ \citep{HI4PI2016}. \citet{Liu2021_AGN} performed spectral analysis for all the sources in the main catalogue, and found that this choice of (average) spectral index leads to an unbiased flux estimation for the whole sample, although a residual uncertainty is caused by the variety of spectral shapes. 

\begin{table}[]
\caption{Sample selection criteria}
     \begin{tabular}{p{0.28\columnwidth}P{0.18\columnwidth}P{0.17\columnwidth}P{0.17\columnwidth}}
    \hline
    \hline
    \multicolumn{4}{c}{Single-band detection [0.2--2.3\,keV]} \\
    \hline
            catalogue & DET\_LIKE & EXT\_LIKE & Sources\\
    \hline
    Full & $\geqslant$5 & $\geqslant$0 & 32684 \\
        Main &    $\geqslant$6 & $\geqslant$0  & 27910 \\
        Supplementary & $<$6 & $\geqslant$0  & 4774 \\
        Point Sources  & $\geqslant$6 & $=$0 & 27369 \\
        Extent-Selected  & $\geqslant$5  & $\geqslant$6 & 542 \\
        \hline\hline
                \multicolumn{4}{c}{Three-band detection} \\
                \hline
        catalogue & DET\_LIKE\_3 & EXT\_LIKE & Sources\\
        \hline
        Hard [2.3-5\,keV] & $\geqslant$10 &  $=$0 & 246 \\
        \hline
    \end{tabular}
    \tablefoot{The single-band detection produces the main X-ray catalogue, together with the supplementary catalogue of faint sources that are less reliable. The point-source catalogue and extended-source catalogues are selected from the main catalogue. The three-band detection results in sources that mostly are already in the single-band detected catalogue, and they are only used to select hard sources focusing on the 2.3-5 keV band.}
    \label{tab:selection}
\end{table}

\subsection{The catalogues}

When a low detection likelihood threshold of $5$ is adopted, the single-band detection results in a large sample of 32684 sources, which includes a relatively high fraction of spurious sources (see \S~\ref{sec:comcon} for details).
Despite the high spurious fraction, we adopted this low threshold because many faint but potentially interesting sources can be detected, possible cases of blended faint sources can be verified by multiple PSF-fitting, \TL{ and faint sources can be effectively masked out when the properties of nearby sources are measured.
As discussed in \citet{Liu2020},} our strategy is to adopt a low threshold in the source detection and to apply further likelihood filtering on the output as needed.
In this paper, we selected the 27910 \TL{single-band detected} sources with a detection likelihood $\ge6$ as constituting the eFEDS Main catalogue. The 4774 sources with detection likelihood $<6$ are kept in a supplementary catalogue, which we also publish here. \TL{The selection criteria for the sub-catalogues are listed in Table~\ref{tab:selection}. The main catalogue contains $27369$ point sources (extent likelihood $=$0), and the full single-band detected catalogue contains $542$ extended sources.}

Figure~\ref{fig:density} shows a \LAST{HEALpix\footnote{http://healpix.sourceforge.net}} map of the eFEDS field, colour-coded with the projected sky density of the eFEDS main catalogue point sources. In the well-exposed part of the field, the catalogue provides a quite uniform source density across the entire covered area (with an average of about 200 sources per deg$^2$). As a general term of reference, this average source density is about 70 times higher than that of the ROSAT All-Sky survey \citep{Boller2016}, about 10 times higher than that of eRASS1, and about a factor 2.5 lower than that of the XMM-XXL survey \citep{Chiappetti2018}.

\TL{
As shown by simulation, the three-band detection is not as efficient as the single-band detection for the main population of X-ray sources, which have soft spectral shapes.
In this work, the only aim of the three-band detection is to select hard sources (particularly AGN) that are detected above 2.3~keV.
We used three bands rather than only the 2.3-5~keV hard band because most of the hard-band observed sources have much stronger signals in the soft band. Including the soft signals improves the measurement accuracy of source position and extent. The source detection likelihood is meanwhile independently calculated in each of the three bands, allowing us to focus only on the hard band.
From the three-band detected sources, we selected} a sample of 246 sources with an extent likelihood of 0 and a 2.3--5\,keV band detection likelihood $\geqslant$10 as the {\it Hard eFEDS} catalogue. 
\TL{The three-band detected sources are not published except for this small {\it Hard} sample.
We adopted a high 2.3--5~keV \texttt{DET\_LIKE} threshold of 10 in order to guarantee a high purity of the sample. Above this threshold lie eight significantly extended sources with \texttt{EXT\_LIKE}$>$47, and all the other $246$ sources have \texttt{EXT\_LIKE}$=$0. 
Only 20 of these 246 sources are not included in the single-band detected main catalogue. The number is small because eROSITA has a small effective area and a high particle background above 2.3~keV. However, we remark that this sample is valuable for AGN demography studies. Although most of them are already in the main catalogue, these sources are detected independently and thus have independent position and fluxe measurements. Most importantly, they have a well-defined selection function, which is different from that of the main sample and is quantified through simulation.}

The content of the catalogues \LAST{is described in detail} in Appendix~\ref{appendix_catalog}. 
Figure~\ref{fig:hist} displays the distributions of fluxes converted from the count rates. The energy conversion factor (ECF; \TL{listed in Table.~\ref{table:bands}}) \FINAL{between} the count rates and fluxes is based on a power-law model with $\Gamma=$2.0 and with Galactic absorption (N$_\mathrm{H}$=$3\times 10^{20}$ cm$^{-2}$).

\begin{figure}[hptb]
\begin{center}
\includegraphics[width=\columnwidth]{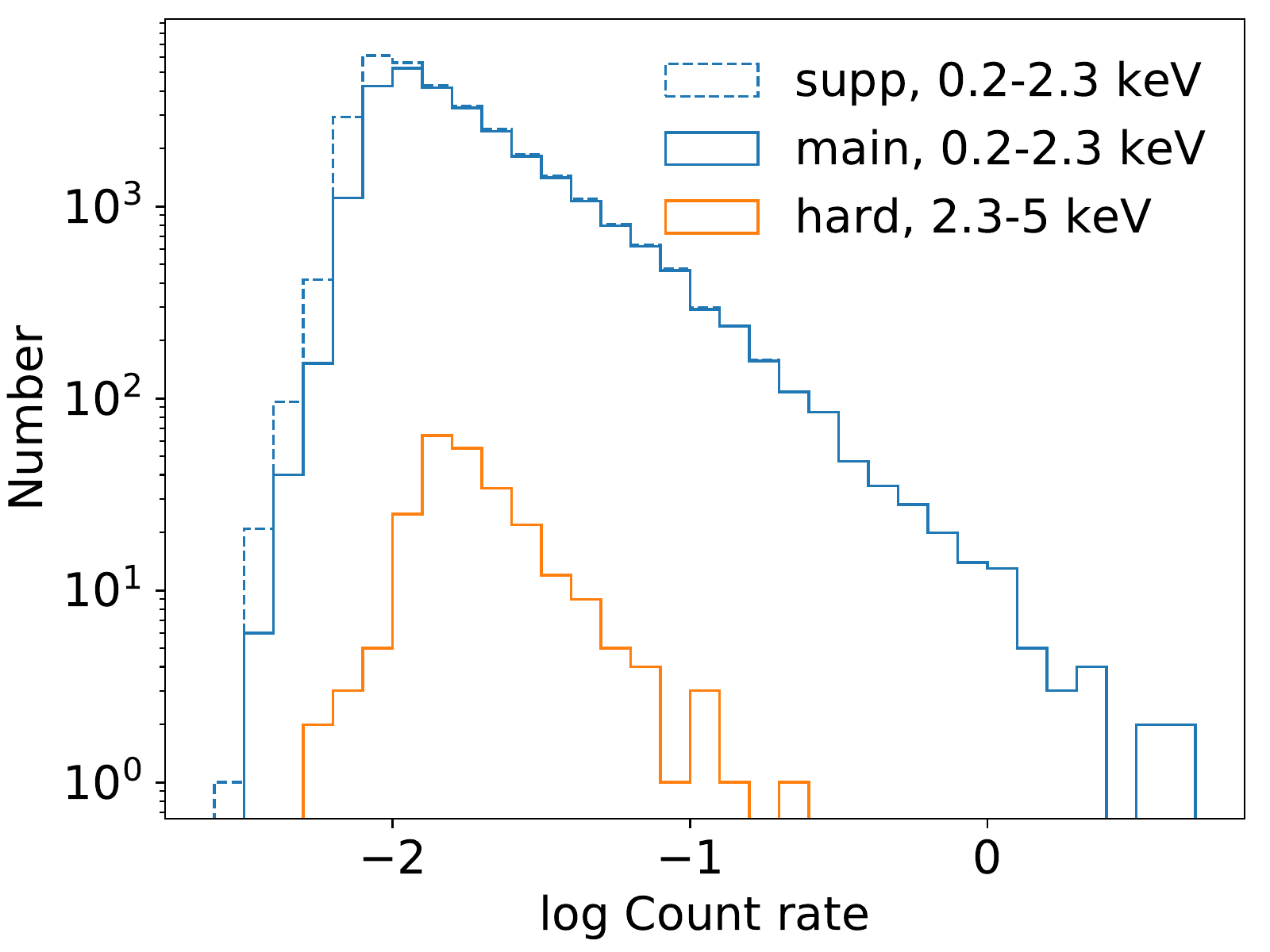}
\includegraphics[width=\columnwidth]{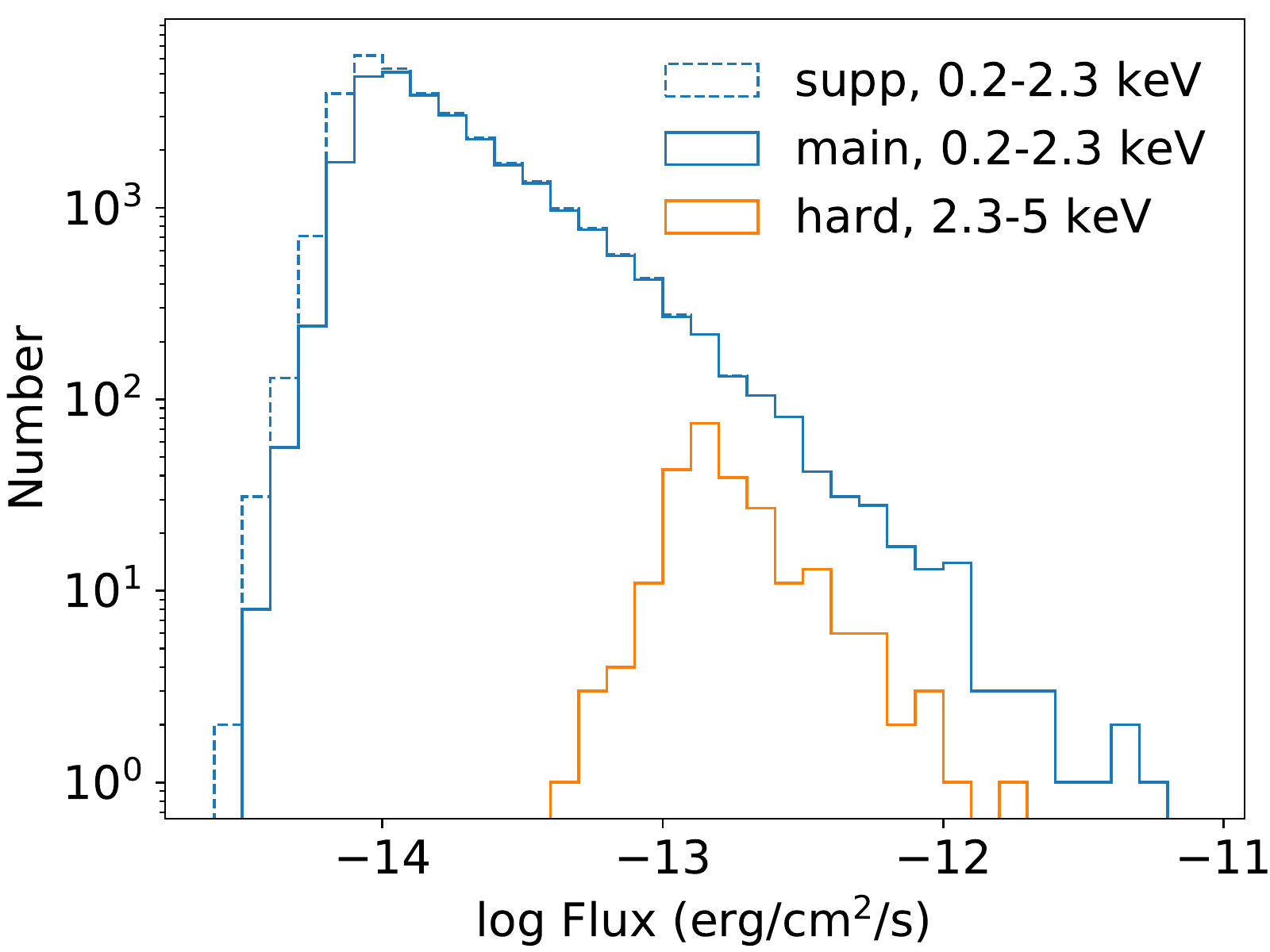}
\caption{Count rate \TL{(counts/s; measured by PSF-fitting) and flux (derived from count rate)} distributions of the main (solid blue) and the supplementary (dashed blue) catalogue in the 0.2--2.3\,keV band and of the hard catalogue (orange) in the 2.3--5\,keV band.}
\label{fig:hist}
\end{center}
\end{figure}

\begin{figure}[hptb]
\begin{center}
\includegraphics[width=\columnwidth]{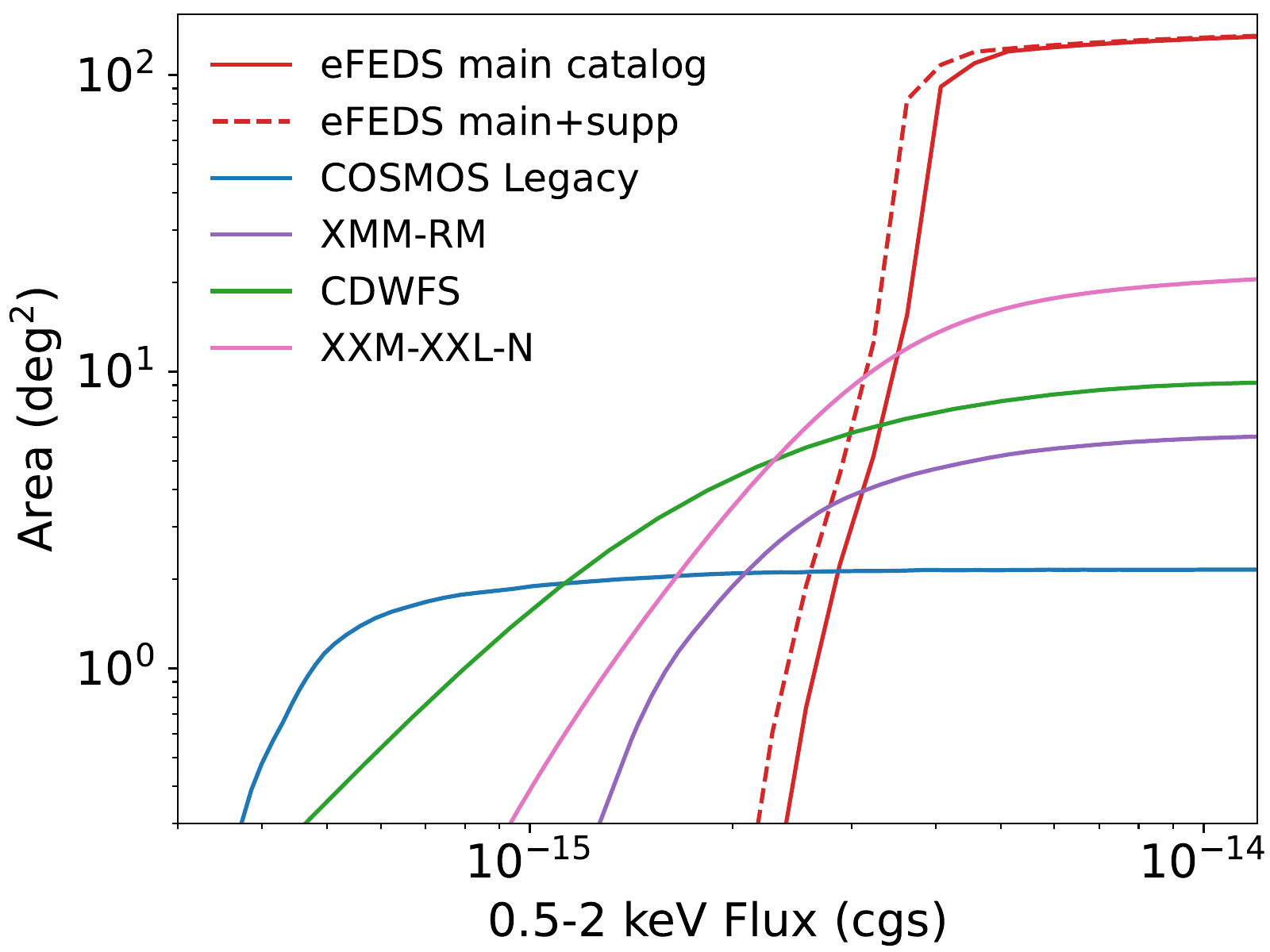}
\caption{eFEDS sky coverage area (red) as a function of 0.5--2\,keV flux calculated using \texttt{ersensmap} at two detection likelihood threshold 5 (for the full single-band sample) and 6 (for the main catalogue). The sky coverage of a few previous contiguous X-ray surveys are plotted for comparison. 
}
\label{fig:sky_cov}
\end{center}
\end{figure}

Using \texttt{ersensmap}, we calculated the sensitivity for point sources in the 0.2--2.3\,keV band.
As a reference, we compare in Figure~\ref{fig:sky_cov}  the sky coverage of the main sample (detection likelihood $\geqslant$6) and of the full single-band detected sample (detection likelihood $\geqslant$5), converted into the 0.5--2\,keV band, with that of a few contiguous \chandra and \xmm surveys, including the XMM-XXL North survey \citep{Liu2016}, the \chandra COSMOS Legacy survey \citep{Civano2016}, the XMM-RM survey \citep{Liu2020}, and the CDWFS survey \citep{Masini2020}.

\begin{figure}[hptb]
\begin{center}
\includegraphics[width=\columnwidth]{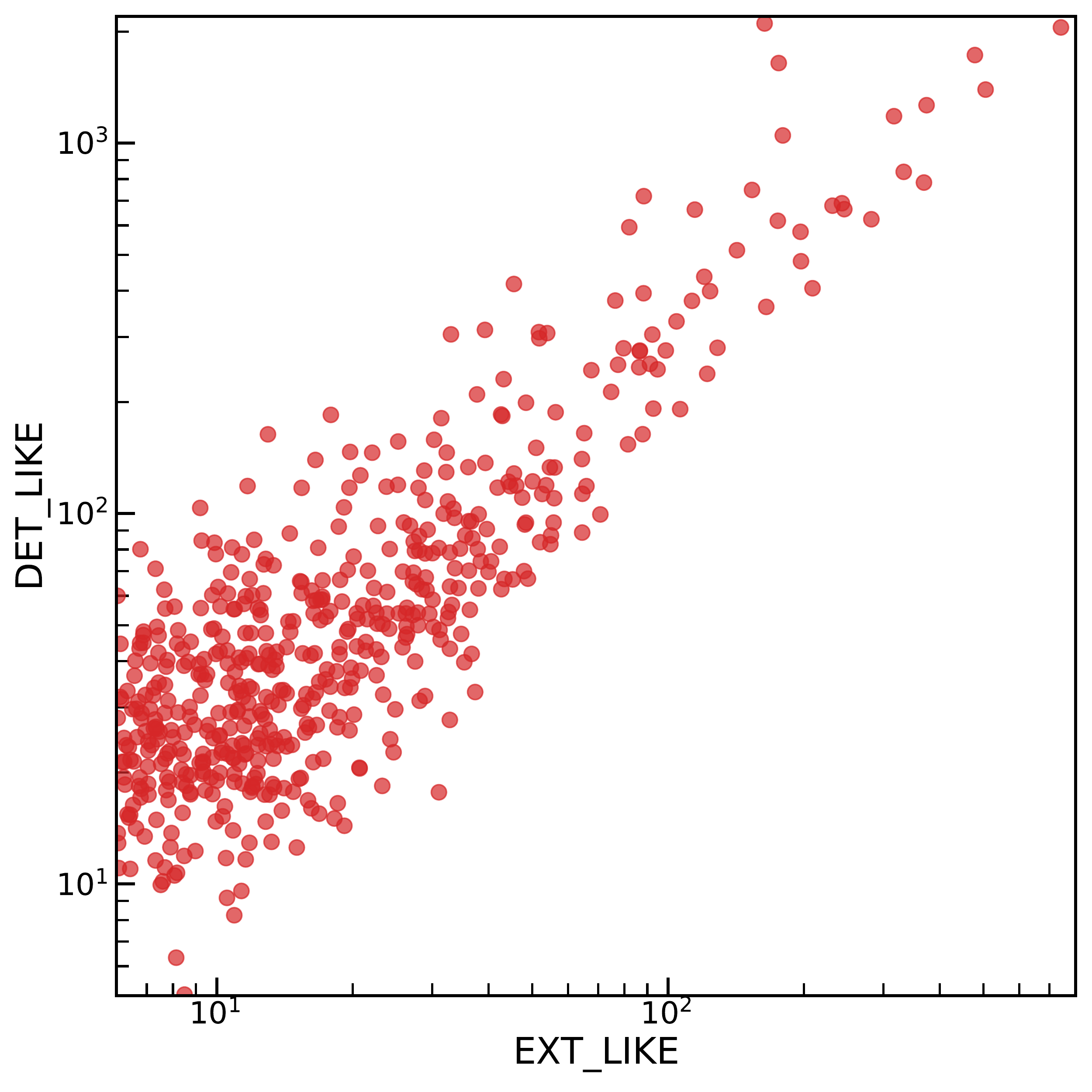}
\caption{Detection likelihood as a function of extent likelihood of 542 extended source candidates in the 
eFEDS catalogue.
}
\label{fig:ext_sample}
\end{center}
\end{figure}

In this catalogue, we detect 542 candidate extended sources with a detection likelihood $\ge$5
and extent likelihood $\ge$6 (see Table~\ref{tab:selection}). This number corresponds to a density of $\text{about four}$ extended sources per square degree over the full eFEDS field. The extent and detection likelihoods of this sample are shown in Figure~\ref{fig:ext_sample}. 
The emission from the detected extended sources was fit using a beta model with a slope fixed to 2/3 and $r_{core}$ as the extent parameter after convolution with the PSF. The fluxes and count rates were then calculated from the normalisation of this fit model and hence correspond to the total integral over the beta model.  During this process, a constant temperature of the intra-cluster medium and a constant value of the Galactic absorption (see above) were assumed. Our source detection algorithm automatically deconvolves the source flux when a nearby point source lies within a radius marked `multrad'. If the multrad parameter is smaller than the physical radius within which the flux of the extended source is measured, emission from the point source might contaminate the signal from the extended source. Furthermore, a diligent analysis of the instrumental background and X-ray foreground has to be performed when the extended source flux is calculated because the source detection algorithm might overestimate the background if the source is very extended. Therefore, the fluxes and count rates given in the main eFEDS catalogue must be treated with extreme care. We provide the corrected count rate, flux, and luminosity measurements of the full extent-selected sample at two fixed radial distances (300~kpc and 500~kpc) in \LAST{\citet{Liu2021_cluster}} and at physical overdensity radius of $R_{500}$ in \LAST{\cite{Bahar2021}}. The extended sources in the sample are characterised in \LAST{\cite{Ghirardini2021b}}, and the optical counterparts are given in \LAST{\cite{Klein2021}}.

\section{Characterisation of the source detection procedure and catalogue properties}
\subsection{Completeness and contamination}
\label{sec:comcon}
\begin{figure}
    \centering
    \includegraphics[width=0.9\columnwidth]{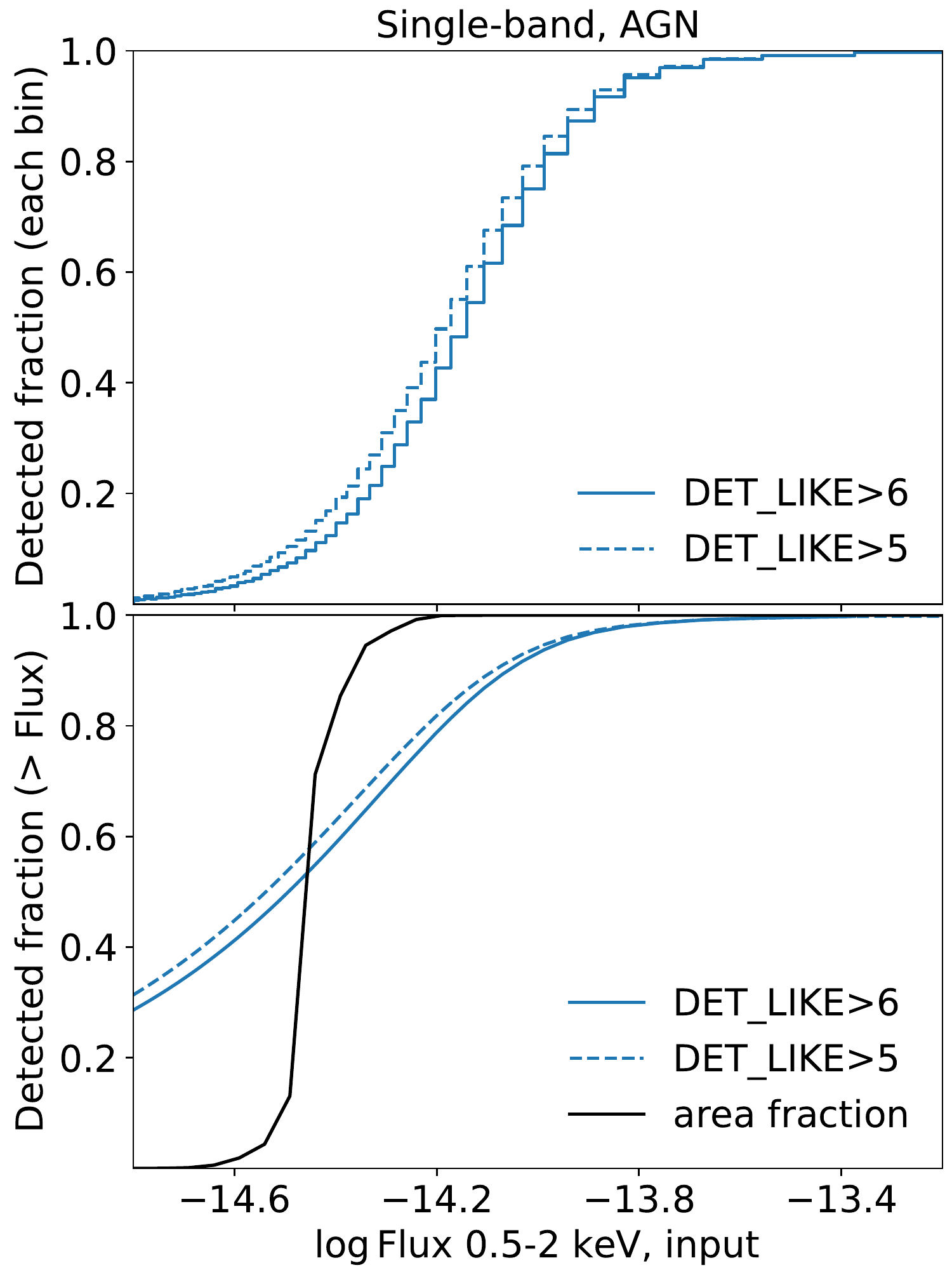}
    \caption{Simulation-measured completeness of AGN in the single-band detected catalogue as a function of the 0.5--2~keV input flux (erg/cm$^2$/s) in differential (upper panel) and cumulative manners. \TL{The solid and dashed blue lines indicate the detected AGN with \texttt{DET\_LIKE}$>$6 and \texttt{DET\_LIKE}$>$5, respectively.} For comparison, we also plot the sky coverage area curve measured by \texttt{ersensmap} adopting \texttt{DET\_LIKE$\geqslant$5} normalised to a total area of 1 (black line).}
    \label{fig:sim_com}
\end{figure}

\begin{figure}
    \centering
    \includegraphics[width=0.9\columnwidth]{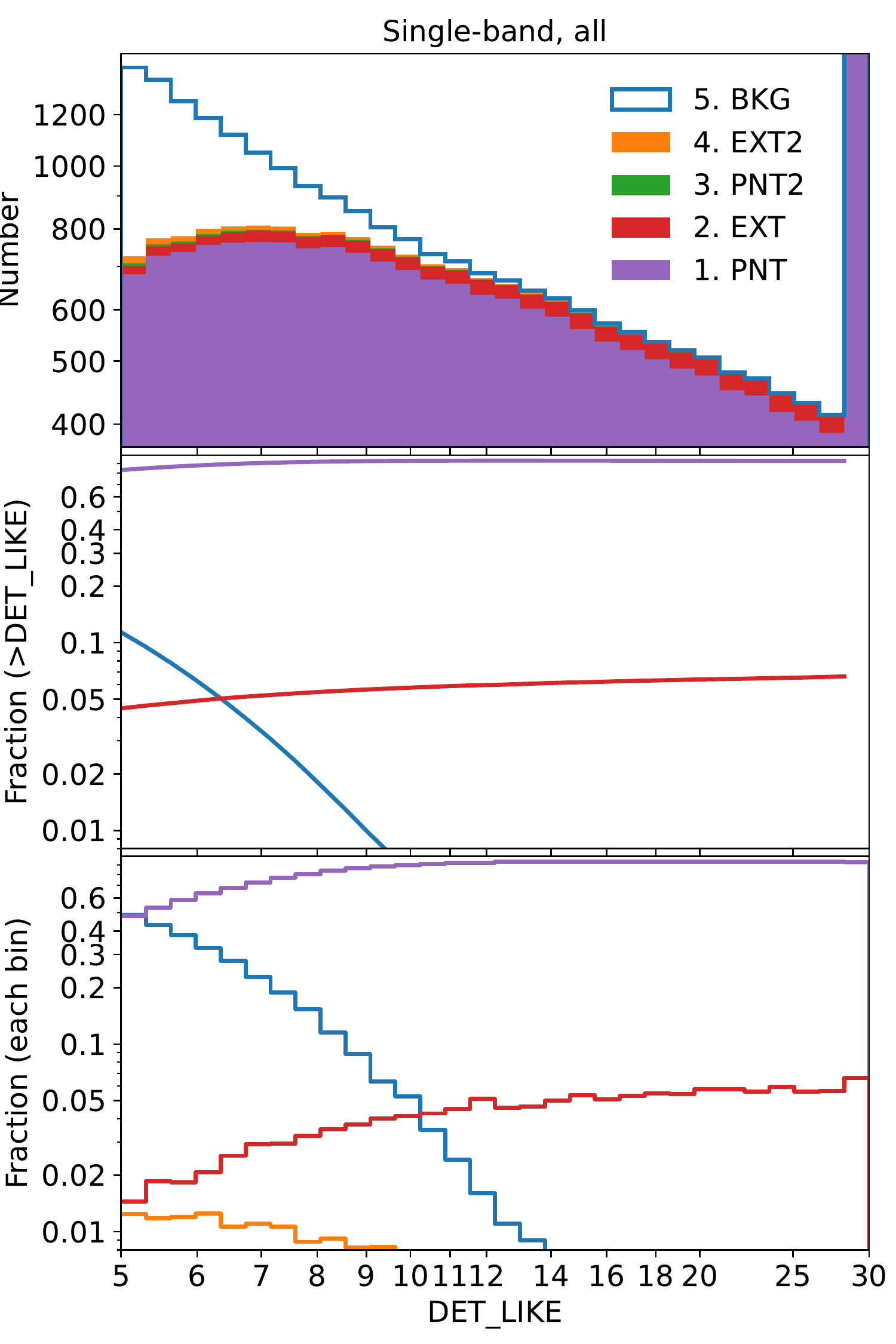}
    \caption{Distributions of all the single-band detected sources as a function of detection likelihood. The top panel displays the histogram of the detected sources. The bottom panel displays the fraction of each class in each bin (differential distributions). The middle panel displays the fraction of each class above a given detection likelihood (cumulative distributions). The colour code for the various classes considered here is as follows: \TL{  purple shows primary counterparts of input point sources, red shows primary counterparts of input clusters, green shows secondary counterparts of input point sources, orange shows secondary counterparts of input clusters, and blue shows spurious sources (background fluctuations).}  \TL{The fractions of green and orange sources are so low that they are not always visible.}
    }
    \label{fig:sim_dis}
\end{figure}

Detection of celestial sources in the low-count regime (as in this case and in most of X-ray astronomy) is a particular \FINAL{realisation} of a stochastic process. Each catalogue generated by such a source detection procedure is statistical in nature, and is always plagued by spurious contaminating sources, as well as by incompleteness or by missed sources.
A high level of completeness (fraction of detected true sources above a give threshold) achieved at a low contamination level (fraction of spurious sources in the catalogue) is the essential figure of merit of any source detection procedure.

We have measured the completeness and contamination of the eFEDS catalogue through an extensive series of simulations \citep[][summarised in Appendix~\ref{sec:simulation}]{Liu2021_sim}. We briefly describe the main outcome of this analysis here.

Fig.~\ref{fig:sim_com} displays the detected fraction of input point sources (AGN and stars) in a differential (top panel) and cumulative (bottom panel) manner. \TL{At a threshold of \texttt{DET\_LIKE}$>5$ and \texttt{DET\_LIKE}$>$6,  94\% and 93\%, respectively, of the simulated point sources} are detected down to a 0.5--2\,keV flux limit of $10^{-14}$ erg cm$^{-2}$ s$^{-1}$. Down to a flux limit of $4\times10^{-15}$ erg cm$^{-2}$ s$^{-1}$ \LAST{, the} completeness reduces to \TL{63\% for \texttt{DET\_LIKE}$>$5 and 59\% for \texttt{DET\_LIKE}$>$6. To guarantee a 80\% completeness, the flux limit is $6\times10^{-15}$ erg s$^{-1}$ cm$^{-2}$ for \texttt{DET\_LIKE}$>$5 and $6.5\times10^{-15}$ erg s$^{-1}$ cm$^{-2}$ for \texttt{DET\_LIKE}$>$6.}
Using \texttt{ersensmap}, we calculated the flux limit corresponding to the detection likelihood threshold (\texttt{DET\_LIKE$=5$}) and thus the sky coverage area curve as a function of this flux limit. When this curve is normalised to a total area of 1 (solid black line in Fig.~\ref{fig:sim_com}), this function also predicts the detectable fraction. However, this fraction only reflects the  \texttt{ersensmap} definition of the detectable flux limit. As displayed in Fig.~\ref{fig:sim_com}, a source above the detectable flux limit might still be missed because of the fluctuation and measurement uncertainty in the source and background or because of blending with nearby sources. A source below the limit still has a significant probability of being detected because of fluctuation.

\TL{Fig.~\ref{fig:sim_dis} displays the distribution of all the single-band detected sources from the simulations as a function of their detection likelihood. The detected sources are divided into five classes: the primary counterpart of an input point source (class 1, purple in Fig.~\ref{fig:sim_dis}), the primary counterpart of an input cluster (class 2, red), the secondary counterpart of an input point source (class 3, green), or an input cluster (class 4, orange),} and spurious sources due to background fluctuations (class 5, blue).
When one input source results in multiple detected sources, the \TL{source with} the largest number of photons is considered as the primary counterpart (class 1 or 2), and the secondary sources (class 3 or 4) correspond to the signal in the outer wing of an input point sources or to substructures or fluctuations in an input cluster. \TL{More details are discussed in \citet{Liu2021_sim}.}

According to the definition, \TL{the detection likelihood \texttt{DET\_LIKE} corresponds to a probability of $ \exp(-\texttt{DET\_LIKE})$ of} one source being spurious. This probability is too low to be plotted in the bottom panel of Fig.~\ref{fig:sim_dis}; the actual spurious fraction (blue line) is significantly higher. This is because the likelihood is defined in an ideal situation, considering only Poissonian fluctuations. In reality, additional uncertainties such as background measurement or source deblending affect the source detection process in every step. The only way to measure the spurious fraction of a catalogue reliably therefore is through detailed simulation, as we have done here.

The middle panel of Fig.~\ref{fig:sim_dis} shows that above a detection likelihood of 5, the sample includes 11.5\% spurious sources. When a likelihood threshold of 6 or 8 is adopted, the spurious fraction reduces to 6.3\% or 1.8\%, respectively.
The availability of a classification for all detected sources means that the simulation provides the fraction of any type of input in any specifically selected sample. In the suite of accompanying papers, we have made extensive use of this valuable information in order to estimate, for example, the fraction of spurious sources in the hard-band selected eFEDS point source catalogue \LAST{(Nandra et al., in prep.)} and the fraction of true/spurious clusters in the extended source catalogue \citep{Liu2021_cluster,Klein2021}, and so on \citep[see more detailed discussions in][]{Liu2021_sim}.

\subsection{Aperture photometry and number counts}\label{sec:lognlogs}
\begin{figure}
    \centering
    \includegraphics[width=\columnwidth]{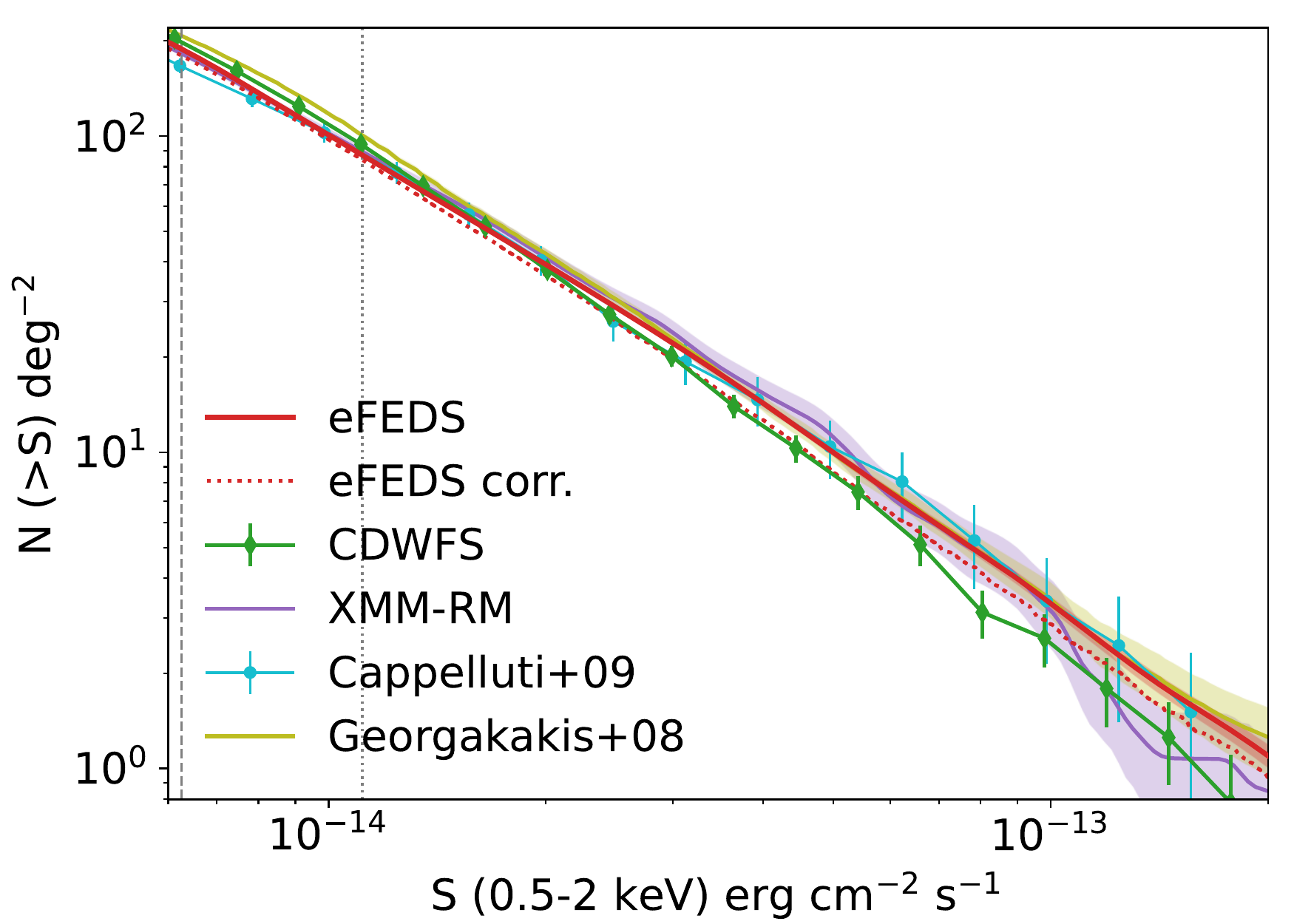}
      \includegraphics[width=\columnwidth]{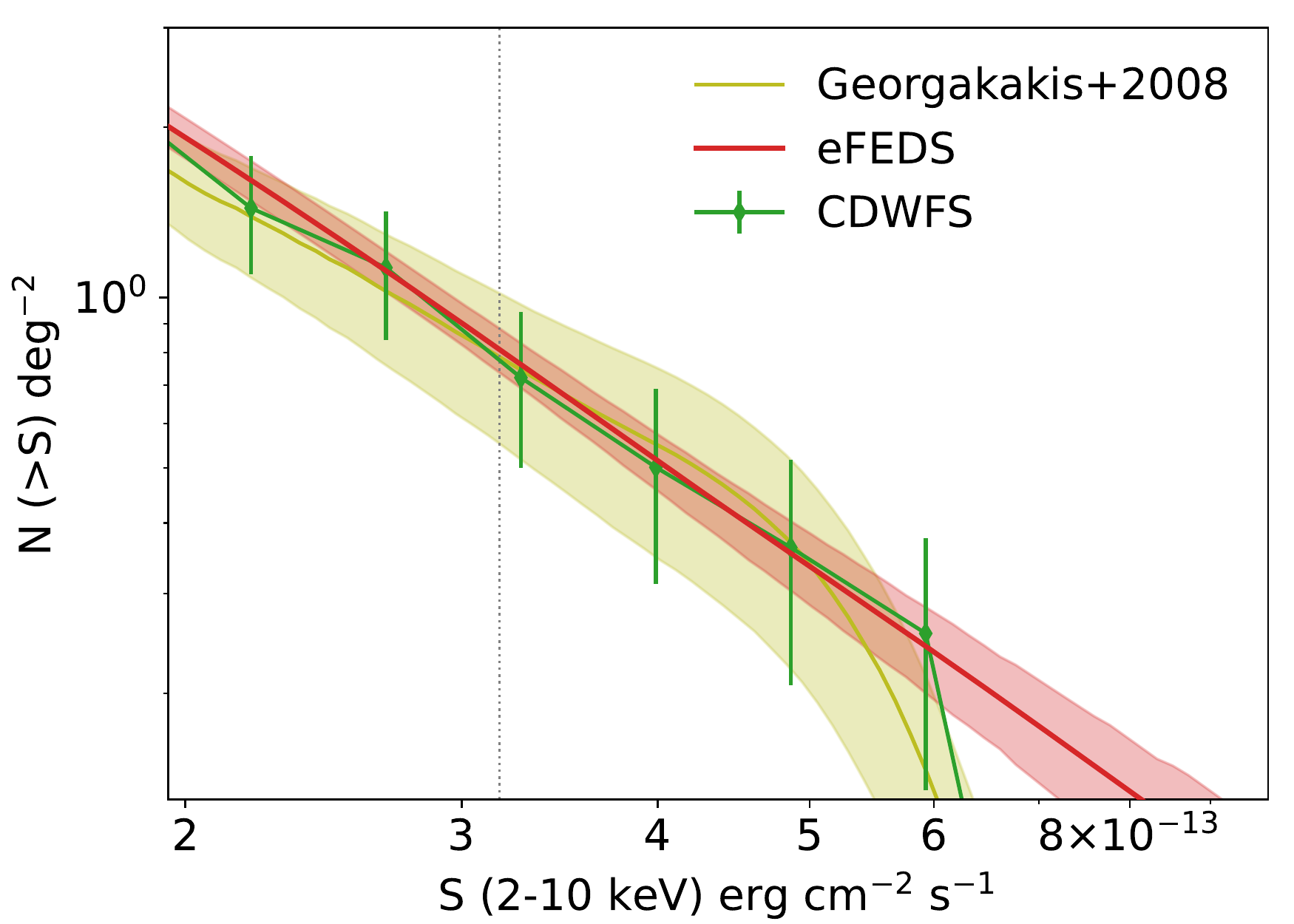}
     
    \caption{Point source number counts as a function of fluxes, corrected for Galactic absorption, for the 0.5-2\,keV (upper panel) and 2-10\,keV (\HB{lower} panel) bands.  The 1$\sigma$ uncertainties are estimated as the square root of the sources number. In both panels, the eFEDS curves are compared to those of \cite{Georgakakis08} (compiled from a number of \chandra survey fields) and \cite{Masini2020} (from CDWFS). In the soft band, we also plot the results from the XMM-COSMOS survey \citep{Cappelluti2009} and from the XMM-RM survey  \citep{Liu2020}. In both panels, the dashed and dotted vertical lines indicate the flux limits corresponding to 10\% and 50\% sensitivity, respectively.
    \TL{In the upper panel, the dotted red line displays the number counts derived by applying corrections according to the simulation results to the distribution of the 0.5-2~keV fluxes measured by forced PSF-fitting.} 
    }
    \label{fig:logNlogS}
\end{figure}
This section presents the X-ray point-source number count distribution of the eFEDS survey, which requires a knowledge of the selection function, that is, the probability of a source with a given flux to be detected. It has been demonstrated that the selection function can be estimated on the basis of aperture photometry \citep{Georgakakis08,Lehmer2012}. 
Therefore, we ran aperture photometry using the \texttt{apetool} task (see Appendix~\ref{app:photometry}). \texttt{apetool} uses the aperture source and background counts to calculate a Poisson false rate for each source, that is, the probability that the source is generated by background fluctuations. This can be expressed in terms of logarithmic likelihood $L$ as $-\ln(probability)$.
This likelihood can be used for the sample selection. It can also be converted into a sensitivity map containing the minimum required number of photons to reach a given likelihood.
This map can then be used to correct for the incompleteness of the eFEDS point-source catalogue as a function
of X-ray flux.
During the forced PSF-fitting photometry for each band, we also created a background map and a source map (source extent model convolved with PSF) in addition to \FINAL{calculating} the fluxes.
Making use of them, we ran \texttt{apetool} within a radius of \FINAL{60\% encircled energy fraction} (EEF), adopting an aperture likelihood threshold of 12. The aperture size and likelihood threshold were selected according to simulation tests \citep{Liu2021_sim}.
In this analysis, we excluded the border of the field, where the exposure is much lower than the typical depth of eFEDS. We adopted the inner region, in which the 0.2--2.3\,keV vignetted exposure value is above 500~s. This region comprises 90\% of the total area.

\begin{figure}
\includegraphics[width=\columnwidth]{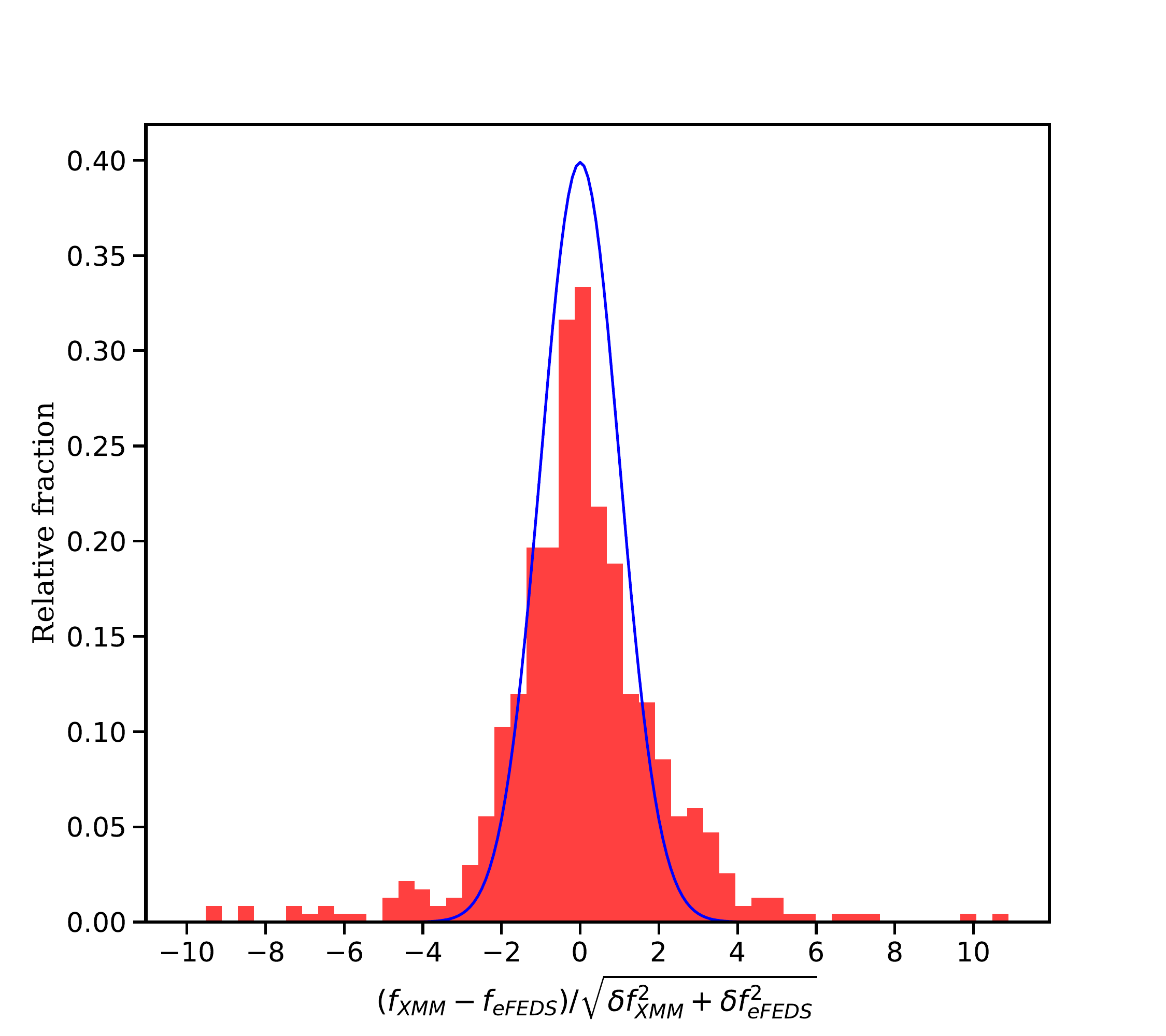}
   \caption{Distribution of the flux difference of X-ray sources between the  XMM-ATLAS and eFEDS observations normalised by the corresponding flux uncertainties added in quadrature. The blue line shows a Gaussian distribution with a mean of zero and a scatter of unity.}
    \label{fig:age-ero-atlas-det}
\end{figure}

Based on the aperture photometry results, we converted the 0.5--2\,keV count rate into flux corrected for Galactic absorption using an ECF of 
$1.104 \times 10^{12}$ cm$^2$/erg, and converted the 2.3--5\,keV count rate into 2-10 keV flux corrected for Galactic absorption using an ECF of 
$5.518\times 10^{10}$ cm$^2$/erg, assuming the same spectrum model as used in \S~\ref{sec:forced_phot}, that is, a power law with a photon index of 2.0. We selected point sources with an aperture Poissonian likelihood $>12$, which resulted in $13457$ sources in the 0.5--2\,keV band and $151$ sources in the 2.3--5\,keV band. 
Based on these two \FINAL{sub-samples} and using the method described in \citet{Georgakakis08}, we calculated the number counts in the \HB{0.5--2} and 2--10\,keV bands and compared them with those measured in \chandra surveys \citep{Georgakakis08} and in the XMM-COSMOS survey (Cappelluti et al. 2009) in Fig.~\ref{fig:logNlogS}.
They agree well.

\begin{figure*}[hptb]
\begin{center}
\includegraphics[width=1.3\columnwidth]{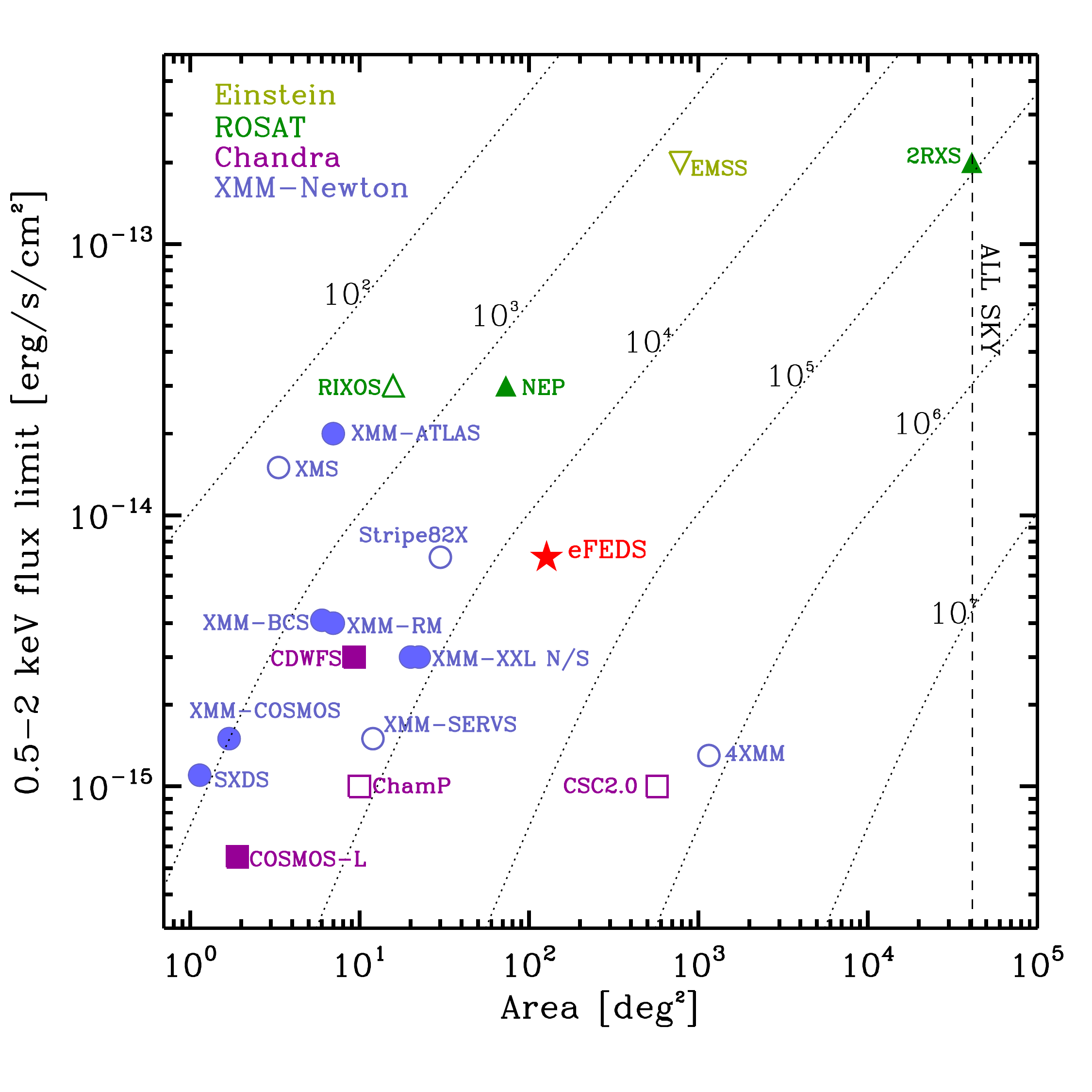}
\caption{Point-sources flux limit (in the 0.5--2 keV energy range) vs. area scatter plot of a few selected X-ray surveys larger than 1 deg$^2$. Existing surveys from Einstein (light green downward triangle), ROSAT (dark green upward triangles), \xmm (blue circles), and \chandra (purple squares) are shown for reference. Filled points mark contiguous surveys, and empty points show non-contiguous ones. To allow a fair comparison, we considered the total area that was covered (x-axis) for each survey and a flux limit that corresponds roughly to a completeness level of 66\%. The dotted lines mark the loci of constant source numbers based on the number counts in the 0.5--2 keV energy rage of \cite{Mateos2008} (double power-law model). The surveys from left to right are The Subaru/\xmm Deep Survey \citep[SXDS; ][]{Ueda2008}; the XMM-COSMOS survey \citep{Cappelluti2009}; the COSMOS-Legacy survey \citep{Civano2016}; the \xmm Medium Survey \citep[XMS; ][]{Barcons2003}; the \xmm Bright Survey \citep[XMM-BCS; ][]{Dellaceca2004}; the XMM-RM survey \citep{Liu2020}; the XMM-ATLAS survey \citep{Ranalli2015}; the Chandra Deep Wide Field Survey \citep[CDWFS; ][]{Masini2020}; the Chandra Multiwavelength Project \citep[ChamP; ][]{Kim2007}; XMM-SERVS \citep{Brandt2020}; the ROSAT International X-ray/Optical Survey \citep[RIXOS; ][]{Mason2000}; XMM-XXL N \& S \citep{Chiappetti2018}; Stripe82X \citep{Lamassa2016}; NEP \citep{Henry2006}; the second Chandra Serendipitous Sources catalogue (CSC2.0; Civano, priv. comm.); the Einstein Observatory Extended Medium-Sensitivity Survey \citep[EMSS; ][]{Gioia1990}; the fourth \xmm serendipitous source catalogue \citep[4XMM; ][]{Webb2020}, and the second ROSAT all-sky survey catalogue \citep[2RXS][]{Boller2016}.}
\label{fig:sensitivity}
\end{center}
\end{figure*}

\TL{\citet{Liu2021_sim} provided another way of measuring the point source number counts based on the detailed eFEDS simulation. The flux distribution of the output AGN catalogue is different from that of the input AGN catalogue mainly because of (i) sample incompleteness (Fig.~\ref{fig:sim_com}), (ii) sample contamination (Fig.~\ref{fig:sim_dis}), (iii) flux overestimation caused by source blending, and (iv) Eddington bias. The ratio of the input and output flux distributions of the simulation can be used to convert the flux distribution of the real catalogue into the intrinsic number counts. The soft-band number counts derived in this way from the real point sources with \texttt{DET\_LIKE}$>$8 are displayed in Fig.~\ref{fig:logNlogS}. This is slightly lower than the number counts measured using the method based on aperture photometry because the simulation-based method has corrected for the effects of sample contamination and source blending, which can only be quantified through simulations. These two effects are negligible in deep survey with high spatial resolution such as the CDWFS. The soft-band CDWFS number counts are slightly lower than those of the other surveys at high fluxes probably for this reason.}

The \texttt{apetool}-generated catalogues were also used to assess the photometric calibration of the eFEDS field by comparing the fluxes of the detected sources to external catalogues. The choice of aperture photometry for this application enables the statistically robust estimation of the random (shot-noise) and \TL{systematic} uncertainties (e.g. Eddington bias) affecting source fluxes \citep[e.g.][]{Laird2009}. This  means that observational effects can be fully accounted for when fluxes from different experiments are compared to test cross-calibration issues. We also chose to use fluxes in the 0.5--2\,keV energy interval because in this standard band, many X-ray catalogues in the literature report fluxes. 

The external dataset adopted in this work is the XMM-ATLAS  survey \citep{Ranalli2015}. This is one of the wide-area ($\rm 6\,deg^2$) and shallow ($F_{\rm 0.5-2\,keV} \approx \rm 2 \times 10^{-14}\,erg\,s^{-1}\,cm^{-2}$) surveys carried out by \xmm and it also overlaps with the eFEDS field. A custom reduction of the XMM-ATLAS survey field was used based on the methods described by \cite{Georgakakis2011}. \AGE{The advantage of using a custom analysis rather than the publicly available XMM-ATLAS catalogue is control over systematics (e.g. Eddington bias and conversion factor of counts to flux)}. The identification numbers of the relevant \xmm observations are 0725290101, 0725300101, and 0725310101. They were reduced using the \xmm Science Analysis System (SAS) version 18. Sources were detected independently in three energy intervals, 0.5--2,  2--8,  or 0.5--8\,keV  to  the  Poisson  false -detection  threshold  of $\rm < 4 \times 10^{-6}$. 

The 987 sources detected in the 0.5--2\,keV band to the threshold above are relevant here. They were compared against a total of 985 eFEDS sources that lie within the XMM-ATLAS footprint and have a detection likelihood in the 0.6--2.3\,keV band of \texttt{DET\_LIKE}$>$10. This threshold was chosen to minimise spurious detections while keeping number statistics high. The calculation of fluxes for the eFEDS and XMM-ATLAS fields was based on aperture photometry and used the Bayesian method described by \cite{Laird2009} and \cite{Georgakakis2011}. A description of the basic flux-estimation algorithm is also provided in \LAST{\cite{Boller2021}}. The fluxes in both samples are estimated in the 0.5--2\,keV spectral band assuming a power-law spectral model with $\Gamma=1.4$ that is absorbed by a Galactic column density of $\log N_H/\rm cm^{-2}=20.3$. We emphasise that this is different from the spectral model adopted for the calculation of fluxes in the main eFEDS catalogue. The reason for this is that the XMM-ATLAS reduction adopts $\Gamma=1.4$ for the determination of fluxes. 

The eFEDS and XMM-ATLAS samples have 616 sources in common. They were identified by matching the two catalogues within a radius of 15\arcsec. For the sky density of the \xmm and ATLAS sources, this threshold corresponds to $\ll1$ spurious associations. Figure \ref{fig:age-ero-atlas-det}  compares the 0.5--2\,keV fluxes of the common sources in the two surveys. It plots  the histogram of the flux difference between the eFEDS and XMM-ATLAS normalised to the flux errors (68\% confidence interval) added in quadrature. This distribution is compared with a Gaussian with unity variance and zero mean. There is no evidence for strong systematic offsets in Figure \ref{fig:age-ero-atlas-det} , suggesting an overall good photometric agreement between the XMM-ATLAS and eFEDS data analysis. There are more sources with large normalised flux differences than expected for the normal distribution, however. We attribute this excess power at the wings of the histogram in Figure \ref{fig:age-ero-atlas-det} to the intrinsic flux variability of AGN. This effect is discussed further in \LAST{\cite{Boller2021}}.

\section{Conclusions}

By publishing and documenting the full creation process of the catalogue of X-ray sources detected in the eFEDS field, 
we here complete the verification of the \erosita design performance. We also demonstrate the ability of the instrument as a powerful survey machine for the X-ray sky.

With the exception of the all-sky surveys, the eFEDS itself represents the largest contiguous X-ray field in the soft \TL{X-ray} energy range, as illustrated in Figure~\ref{fig:sensitivity}, where we show the \TL{point-source} flux limit (in the \HB{0.5--2}~keV energy range) -- area scatter plot of X-ray surveys larger than 1~deg$^2$. 
In terms of the sheer number of detected sources excluding all-sky surveys, eFEDS also stands out as the richest contiguous survey field to date. 

This paper serves a twofold purpose: it makes the catalogues of X-ray sources detected in the eFEDS field during the \srg/\erosita PV observations of 2019 public for further scientific investigations, and  it describes the data processing and analysis of \erosita field-scanning observations in a comprehensive manner for the first time, including the relevant dedicated software tools. The tools are described in the appendices.

\begin{acknowledgements}
This work is based on data from eROSITA, the primary instrument aboard SRG, a joint Russian-German science mission supported by the Russian Space Agency (Roskosmos), in the interests of the Russian Academy of Sciences represented by its Space Research Institute (IKI), and the Deutsches Zentrum f\"ur Luft- und Raumfahrt (DLR). The SRG spacecraft was built by Lavochkin Association (NPOL) and its \FINAL{sub-contractors}, and is operated by NPOL with support from the Max Planck Institute for Extraterrestrial Physics (MPE).

The development and construction of the eROSITA X-ray instrument was led by MPE, with contributions from the Dr. Karl Remeis Observatory Bamberg \& ECAP (FAU Erlangen-N\"urnberg), the University of Hamburg Observatory, the Leibniz Institute for Astrophysics Potsdam (AIP), and the Institute for Astronomy and Astrophysics of the University of T\"ubingen, with the support of DLR and the Max Planck Society. The Argelander Institute for Astronomy of the University of Bonn and the Ludwig Maximilians Universit\"at Munich also participated in the science preparation for eROSITA. The eROSITA data shown here were processed using the eSASS software system developed by the German eROSITA consortium.
We thank Alberto Masini for providing us the data for CDWFS to be included in Fig. 8 and 12. N. Clerc acknowledges support by CNES.

\LAST{Some of the results in this paper have been derived using the HEALPix \citep{Gorski2005} package.}
\end{acknowledgements}

\bibliographystyle{aa}
\bibliography{eFEDScatbib}

\begin{appendix}

\section{eSASS data analysis software package}
\label{appendix_esass}

This appendix describes the software tasks comprising the \erosita Science Analysis Software System (eSASS). eSASS provides a set of command-line tools for pipeline processing the \erosita data and for performing interactive data analysis tasks. The eSASS tasks interact with the \erosita calibration database described in Appendix \ref{appendix_caldb}. Calibrated data products provided by the data analysis pipeline are described in Appendix \ref{appendix_products}. A full description of each eSASS task including usage examples is available online\textsuperscript{\ref{webdoc}}.

\subsection{X-ray event processing}

\paragraph{evprep.}
This task generates an event-list file in the format agreed for \erosita from the raw FITS\footnote{\cite{Wells1981}} files created from telemetry by the archiver software. 
\FINAL{The \texttt{evprep} task performs many other corrections, including the detection of a variety of error and out-of-limit conditions; setting of corresponding bits of the event FLAG masks; the correction of offsets in the Camera Electronics (CE) time stamps as specified in the calibration database; the conversion of times in Good Time Intervals (GTI) and Housekeeping (HK) extensions from the low-cadence but reliable ITC time system to the higher-cadence CE time system; unpacking of the compressed event energy data for observations performed in PMENV2 mode; the exclusion of bad time intervals read from the calibration database from the sequence of GTIs; and the calculation and storage of dead time corrections.}


\paragraph{ftfindhotpix.}
The task \texttt{ftfindhotpix} is a tool for detecting and saving the positions of bad pixels by analysing the data with statistical methods. The task can be run in either `single` or  `block` mode, where the Poisson mean value of a single pixel (and its immediate neighbours) or a block of pixels is compared to a given threshold that denotes the maximum probability in percent for a false positive of a bad pixel. In addition, a number of additional parameters such as the number of frame bunches to consider, the minimum number of events in a frame bunch, and the minimum fraction of frames for which is a pixel is considered bad, can be optimised to decrease the probability of false positives. \HB{In the current version of the pipeline,} the detection of bad pixels is turned of,f and \texttt{ftfindhotpix} only sets bits of the FLAG mask for events that lie on or next to a pixel listed in the calibration database as bad (various types are recognised thereof), and writes valid entries from the calibration to a BADPIX extension in the event list file.

\paragraph{pattern.}
The charge cloud released by the absorption of an X--ray photon may extend over several pixels. The task \texttt{pattern} tries to identify these pixels, so that the total released charge can be reconstructed. This is not always possible in a unique way, however (e.g., when the charge clouds released by two photons overlap). The general driver for the photon reconstruction in \texttt{pattern} is to find the simplest, most likely explanation for the observed charge distribution. If no such explanation can be found, the pattern is marked as invalid. Valid patterns consist of 1\,--\,4 pixels, with 1 and 2 pixel patterns (`singles' and `doubles') being the most important ones. The task \texttt{pattern} performs an important role in the whole processing chain because it affects key performance parameters of \erosita, such as the spectral resolution and sensitivity, and also the spatial resolution. 

\paragraph{energy.}
The task \texttt{energy} tries to reconstruct the energy of each detected photon from the charge distributions found in individual pixels and as a byproduct, also constrains its sub-pixel position. It makes use of the results of the \texttt{pattern} task, which must thus be executed beforehand. The \FINAL{reconstruction of the} energy of an incident photon is essentially performed in three steps: First, a charge transfer inefficiency (CTI) correction: the charge loss caused by shifting the charge from its original location to the readout node must be corrected for. Second, a gain correction: the \erosita CCDs are characterised by parallel readout, where each transfer channel is equipped with an amplifier of its own, and the amplification factors need to be individually determined. Third, a recombination of the reconstructed energies from all the pixels of a pattern. In addition to these essential steps for X--ray spectroscopy, it is also possible to obtain an improved location of the centre of the charge cloud from the individual components, so that the location where the X--ray photon had hit the CCD can be determined with sub-pixel resolution \citep{Dennerl2012}. The task \texttt{energy} is crucial for the whole processing chain because the absolute energy scale and the spectral resolution rely on it.

\subsection{Spacecraft attitude and boresight}
\label{app:att}

\paragraph{attprep.}
The task \texttt{attprep} converts the attitude time series from the SRG orientation sensors  (\texttt{SED26/1}, \texttt{SED26/2}, \texttt{BOKZ}, or \texttt{Q-Gyro}, see also Appendix \ref{appendix_caldb}.4) into the eSASS  intermediate-attitude format.  The input attitude is provided as a time series of orientation quaternions for the  \texttt{SED26/1}, \texttt{SED26/2}, and \texttt{Q-Gyro} FITS files and as an orientation matrix for the \texttt{BOKZ} FITS file. The nominal cadence is 1 per second, with time tags given in SRG spacecraft clock (SCC).  The intermediate attitude  describes the orientation of the nominal coordinate system of the central \erosita camera TM1 as a time series of RA (J2000), DEC (J2000), and roll angle, which is defined as the clockwise angle of the camera X-axis with respect to the north direction. The rotational transformations between the sensor coordinate systems and the TM1 camera coordinate system are stored as rotation quaternions in a calibration file for each sensor. The output FITS file contains the intermediate-attitude dataset from the primary sensor in the the first FITS extension and all sensor specific attitude datasets in additional extensions. The primary sensor is the \texttt{Q-Gyro}, if present, or else the \texttt{SED26/1}, \texttt{SED26/2}, or \texttt{BOKZ} in this order. 

\paragraph{telatt.}
The task \texttt{telatt} calculates attitude time series that are specific for one of the telescope modules TM1 to TM7. The input attitude is read from the intermediate-attitude file written by the task  \texttt{attprep}. The camera-specific boresight angles are read  from calibration files in the form of Euler angles and are applied to  the input attitude. The resulting attitude time series is written as RA (J2000), DEC (J2000), and roll angle either to a separate FITS file or as a FITS extension to the camera-specific events file.

\paragraph{evatt.}

The task \texttt{evatt} is used to project event positions onto equatorial sky coordinates. The task reads an event list for a single telescope module, the corresponding attitude time series (as written by task \texttt{telatt}), and a GTI table. For all events in the intervals specified by the GTI table, the telescope attitude is interpolated to the arrival time of each event. The  event position in the detector coordinate system given in the detector pixel columns RAWX, RAWY and sub-pixel information in the SUBX, SUBY columns of the event table is projected onto equatorial coordinates by means of a tangential projection with the interpolated attitude and using the plate scale from a calibration file.

\paragraph{radec2xy.}

The task \texttt{radec2xy} computes X and Y sky pixel coordinates (sine projection; pixel size $0\farcs05$) corresponding to the right ascension and declination (J2000) event coordinates in the \erosita event tables. The projection centre must be specified in the command line. Sky pixel coordinates are required for image binning (\texttt{evtool}) and may be used for spectrum and light-curve extraction (\texttt{srctool}).
\subsection{X-ray event binning}
\label{app:events}
\paragraph{evtool.}

The task \texttt{evtool} provides three capabilities: \FINAL{the merging of several event lists, the filtering of events on a subset of the canonical set of columns, and the creation of images.}

\FINAL{Merging is performed on all canonical extensions of the input files. Superfluous members of sets of identical rows found in the event tables are discarded. The task cannot correctly merge input files that have been subject to differing filterings, or that are inconsistent in a number of other tested ways; warnings are issued if such inconsistencies are detected.}
 
\FINAL{The columns and types of event filtering are given as follows. The \texttt{FLAG} column can be filtered via the specification of a hexadecimal bit mask, for which the filtering sense can be chosen to be either exclusive or inclusive. The \texttt{PAT\_TYP} column is filtered via an integer in the range 0-15, which is interpreted as a 4-bit mask, one bit for each of the accepted patterns. The \texttt{TM\_NR} column is selected via a list of desired numbers. The \texttt{PI} column is filtered via lists of lower and upper energy bounds. The \texttt{TIME} column is filtered via a GTI specification; either directly, or by giving the base name of a GTI extension, or by specifying an external FITS file. Finally, the \texttt{RA} and \texttt{DEC} columns can be filtered via a region specification.}

\FINAL{When \texttt{evtool} is used to create FITS images, the pixel sizes on the sky, the image dimensions in pixels, and the image centre location, are all specifiable. Some auto-sizing operations are also available. Most of the extensions in the input event list(s) are optionally discardable for image output.}

\paragraph{srctool.}
The \texttt{srctool} task is responsible for the generation of standard source products, including spectra, response matrices, background spectra, and light curves.
It takes as input the calibrated event file, source lists, region files, and the calibration data.
The task is designed to create products that take the scanning of the telescope across the sky into account.
The mechanism it uses to do this is to take a set of sample points as a function of position and as a function of energy, given a source model and extraction region.
The source regions and background regions are both sampled in this way.
These samples are propagated through time to take the vignetting and bad pixels as a source scans across the detectors into account, which is included in the computation of the source-specific effective area curves and in the area calculation for light curves.
The GTI and exposure for a source are computed from when the sample points for a source enter and exit the field of view.
Similarly, the area on the sky of the source, when used for background subtraction, and the geometric area of the extraction region are computed as the average values computed from the sample points during the GTIs.
The accuracy to which the output effective areas and exposure times is calculated depends on the parameters giving the spacing of the sample points on the sky and in time.
\texttt{srctool} takes the source morphology into account by a model chosen by the user (e.g. a point source, top hat, beta model, or a provided image).
If PSF losses are taken into account, the source model is convolved by the PSF to compute how much flux is lost outside the extraction region for each time step.
The task supports a number of geometric regions for source extraction, including circles, ellipses, and boxes, in addition to a generic mask image option, all of which can be combined or subtracted.
\texttt{srctool} also has the ability to compute extraction regions automatically given a source list.
In this mode, the aim is to increase the source and background extraction circular regions until the signal-to-noise ratio is maximised, while taking into account excluding neighbouring sources.

\subsection{Map creation}

\paragraph{expmap.}
The \texttt{expmap} task generates FITS-format exposure maps to match the location and dimensions of a supplied template image in register. The exposure is expressed in units of the time (seconds) that each sky location was in the field of view. It can optionally be folded with the telescope vignetting function in each energy band. The algorithm samples periods of GTI in the input event list and projects the CCD onto the sky at each sample time according to the record of spacecraft attitude contained in \texttt{CORRATT} extensions. Other inputs to the maps are bad pixels as listed in the \texttt{BADPIX} extensions, dead time as recorded in the \texttt{DEADCOR} extensions, the detector mask, and the vignetting function. While the vignetting function is energy dependent, we followed the approach of \xmm exposure maps (XMMSAS\footnote{{\em Users Guide to the XMM-Newton Science Analysis System}, Issue 16.0, 2021 (ESA: \xmm SOC)}) to not weight the vignetting with an assumed spectral model. This results in systematic errors of the vignetted exposure and derived count rates and fluxes in the energy bands of interest in the few-percent range. Future versions of the \texttt{expmap} task \HB{will} provide an option to fold the vignetting function with a suitable spectral model. Exposure maps for individual telescope modules and also weighted all-\erosita `merged' maps are available as outputs.

\paragraph{ermask.}
The task \texttt{ermask} uses the exposure maps to calculate detection masks.
The masks are FITS images with the same dimensions as the exposure maps. 
Pixels in image areas that can be used for source detection are set to 1, all other pixels are set to 0. The area selection is based on two configurable thresholds \FINAL{setting either a lower limit on the exposure as a fraction of the maximum map value or an upper limit for the spatial gradient of the exposure map. In the current pipeline versions only the fractional exposure threshold is set.}

\paragraph{erbackmap.}
The task \texttt{erbackmap} calculates a background map based on a photon count image, the corresponding exposure map, the detection mask, and an initial source list.
In a first step, the task calculates a mask to blank out circular regions around
the sources from the input list. The radii of these circles depend on the source count rate, the PSF, and on source extent parameters from the input list. A second step smooths the remaining images and interpolates the map into the masked out regions around the input sources.
For this step \FINAL{either a two dimensional smoothing spline or an adaptive smoothing algorithm (recommended) can be selected.}
The adaptive smoothing algorithm convolves the input images with Gaussian 
smoothing kernels of different scales and chooses the smoothed map for each pixel that reaches the user-defined signal-to-noise ratio.

\paragraph{ersensmap.}
The task \texttt{ersensmap}  uses the eSASS exposure maps and background maps to estimate the detection limits in \erosita observations. For each map pixel, the task calculates  the detection limits for aperture methods as employed by the task \texttt{erbox} or for the PSF-fitting method as used by task \texttt{ermldet}. In the aperture case, the algorithm iteratively determines the source flux necessary to reach the given likelihood threshold according to equation~\ref{eq:pgamma}. In the PSF-fitting case, the task calculates the flux necessary to reach the likelihood threshold determined by the $C$ statistic (equation~\ref{eq:cash}). In both modes, the task can also handle the case where several input images are used for simultaneous source detection as implemented  in \texttt{erbox} and \texttt{ermldet}. For the conversion between fluxes and count rates,  energy conversion factors have to be given as task parameters.

\subsection{Source detection and photometry}
\label{app:photometry}

\paragraph{flaregti.}
The task \texttt{flaregti} creates a set of GTIs that optimise the ability of sources to be detected or created from a fixed count rate threshold.
The first step is to remove bright point sources using a simple detection algorithm that identifies bright pixels in a binned image, which are later excluded from the light-curve production.
A light curve of the count rate per unit area is then computed from the data in a given energy band, given the input GTIs.
A regular grid of points is chosen over the area of the sky that is contained within the event file.
\JS{For each grid point, light curves are constructed by selecting a subset of time bins from the total point-source masked light  curve for those periods when the grid point in question is within the field of view.}
Each of these lightcurves is analysed to either choose an optimal threshold for that spatial location in standard operation, or to apply a fixed threshold.
When an optimal threshold is calculated, the task chooses a threshold that minimizes the flux at which a source of a specified size can be detected against the average sky surface brightness.
When it has a set of thresholds for each grid point, the task then identifies for each time bin within the light curve the grid point that is closest to the centre of the field of view.
For each time bin, the threshold of this nearest grid point and the measured rate is used to decide whether a time bin is good or bad, and the tool then constructs the GTIs.
These flare-based GTIs are merged with the original input GTIs to make new combined GTIs for each TM.
The task then repeats the whole processes back to the source detection stage based on these new GTIs for a given number of iterations.
\texttt{flaregti} writes the output FLAREGTI GTI data back into the original event file as extensions or to a separate GTI file, and optionally writes light curve, point-source mask, and spatial threshold files.

\paragraph{erbox.}
The task \texttt{erbox} is based on a sliding-box algorithm to detect peaks 
in the input count images.  \FINAL{The task can either be run in local mode, which estimates the background from a region surrounding the  source box, or
in map mode, which uses the background maps generated by \texttt{erbackmap} at the position of the source.}
%
%
The algorithm first smooths the input images with a beta-function-type kernel and then searches for peaks in the smoothed images. This peak search can be repeated several times after rebinning the input images by  a factor of $2 \times 2$, which effectively doubles the box size. The peaks are then tested for statistical significance by comparing  the number of box counts $n_i$ with the expected background counts $b_i$ in each input image $i$ using the (logarithmic) likelihood 

\begin{equation}
\label{eq:pgamma}
 L_i = -\ln P_\Gamma (n_i,b_i) 
,\end{equation}
where $P_\Gamma$ is the regularised incomplete Gamma function 

\[ P_\Gamma (a,x) = \frac{\int_0^x e^{-t} t^{a-1} dt }{\int_0^\infty e^{-t} t^{a-1} dt }. \]
In the case of multiple input images (e.g. in different energy bands), a combined likelihood is computed using Fisher's method \citep{Fisher32}. The probability values $P_i$ from $n$ independent tests of the same null hypothesis can be combined as  \( L'=-2 \sum_{i=1}^n \ln P_i \), which follows a $\chi^2$ distribution with $2n$ degrees of freedom. The combined detection likelihood of a source can therefore be calculated as

\[ L_\mathrm{det} = -\ln \left(1 - P_\Gamma \left( n, \sum_{i=1}^n L_i \right )\right). \]
The source list is filtered using a threshold on the combined likelihood. For the significant sources, the following parameters are calculated for each input image and for the combined dataset:

\begin{itemize}
    \item raw box counts,
    \item PSF-corrected box counts with errors,
    \item PSF-corrected count rates with errors,
    \item source fluxes with errors,
    \item background counts,
    \item detection likelihood values,
    \item source exposure values (for each input image only),
    \item centroid positions in image and equatorial coordinates (combined values only) with errors,
    \item the image rebinning step in which the source is most significant,
    \item hardness ratios of the form $HR_{12}= (cr_2-cr_1)/(cr_1+cr_2)$ with errors, \GL{where $cr_1$ and $cr_2$ are the count rates in two of the input energy bands}.
\end{itemize}

\paragraph{ermldet.}

The task \texttt{ermldet} applies a PSF-fitting algorithm to determine source parameters for a list of input positions. It can also apply multi-PSF fits in order to deblend neighbouring X-ray sources. The source PSF can be generated using the following methods:

\begin{itemize}
    \item from  2D PSF images stored in calibration files for a grid of photon energies and off-axis angles (pointed observations).
    \item from 2D images of the PSF averaged over the detector area and stored for a grid of photon energies.   
    \item reconstruction of a source-specific PSF by averaging the PSFs of each source photon, the event-specific PSFs are stored in calibration files as sets of shapelet coefficients for a grid of energies and detector positions.
    \item calculation of event specific PSFs using the shapelet coefficients stored in calibration files ("photon mode").
\end{itemize} 

A maximum likelihood fitting procedure is applied to the sources in the input list, starting with the most significant source and then working on the sources in the order of descending likelihood. Neighbouring sources can be combined for simultaneous fitting, and  input positions can also be split into multiple sources.

The source models are generated using  PSFs that are optionally folded with an extent model.
The extent model is either a Gaussian kernel or a beta model of the form
\[ f(x,y)= \left(  1+\frac{(x-x_0)^2 + (y-y_0)^2}{r_c^2 }  \right)^{-3\beta+1/2} \]
with $\beta=2/3$ . The core radius $r_c$ is a free fit parameter. 
 The models are  multiplied with the exposure map to account for instrumental effects, and the background is added.
The model parameters are

\begin{itemize}
    \item $x,y$ positions in image coordinates for each source,
    \item source extent radius for each source,
    \item source count rate for each source and input image.
\end{itemize}

The source parameters are optimised by minimising the C-statistic \citep{Cash1979},

\begin{equation}
 \label{eq:cash}
  C = 2 \sum_{i=1}^N (e_i - n_i \ln e_i )
,\end{equation}
where $e_i$ is the expected model value of pixel $i$, $n_i$ is number of photon events in pixel $i$, and $N$ is the total number of image pixels used for the fit.
When the fit has converged at a set of best-fitting parameters, the significance of each source is tested  by calculating
\[ \Delta C = C_\mathrm{null} - C_\mathrm{best}, \]
where $C_\mathrm{null}$ is the $C$-statistic of the null hypothesis (i.e. model with zero net counts) and $C_\mathrm{best}$ is the value for the best-fitting model. The fitting algorithm of \texttt{ermldet} can optionally obtain an individual PSF for each photon event depending on its energy and detector position; where applicable, the event PSF is convolved with the extent model.
\GL{The $C$-statistic is then calculated according to equation \ref{eq:cash} , where $e_i$ is now the event-specific model value
in the image pixel in which the event  was detected, and $C$ is summed
over the  $N$ events with index $i$ in the source region using $n_i=1$.
}
According to \cite{Cash1979}, $\Delta C$  under certain conditions approximately follows a $\chi^2$ distribution with the number of degrees of freedom $\nu$ equal to the number of free model parameters. Hence the probability $P$ that $\Delta C$ results from a chance fluctuation of the background can be calculated using the regularised incomplete Gamma function $P_\Gamma$,

\[ P = 1 - P_\Gamma \left(\frac{\nu}{2},\frac{\Delta C}{2}   \right)\,. \]
Based on the value of $P$, a logarithmic detection likelihood 

\[ L = -\ln(P) \]
is obtained for each source. In case of simultaneous multi-source fits, sources falling below the user-defined likelihood threshold are removed and the fitting procedure is iteratively repeated.
The conditions for $\Delta C$ to follow a $\chi^2$ distribution are not completely fulfilled in the low count rate regime, and thus $P$ does not correspond to the actual false-alarm probability. We used extensive simulations \citep{Liu2021_sim} to determine the false-detection probabilities at various levels of $L$.
\GL{An extent likelihood quantifying the probability that an extended model is required to fit the count distribution of a source is calculated in a similar fashion. In this case, the 
value $C_\mathrm{null}$ for the null hypothesis is calculated for the best-fitting point source model. Again, the relation between the extent likelihood $L_\mathrm{ext}$ and the incidence of false classifications as extended sources needs to be calibrated by simulations.
If the extended model fit results in an extent likelihood below the user-defined threshold, the fit is repeated  with a  PSF model and
the fit parameters of the point source model are written to the source list. } 
After each source fit, the best-fitting model is added to the internal background map used for the model fits of subsequent (fainter) sources. The final source model $+$ background maps can be written as FITS images.

Assuming $\Delta C$ follows a $\chi^2$ distribution, we can calculate the parameter errors corresponding to $68\%$ confidence intervals by varying the parameter of interest from its best value in both directions until $C=C_\mathrm{best} + 1.0$ is reached.  
These boundaries are determined by an iterative procedure, and the  errors for both directions are averaged and written to the output table.
When for one direction the algorithm fails to converge after the given number of iterations, the errors determined for the other direction are adopted.
When  no error margin can be determined in either direction, a zero 
is written to the output table.
The positional errors are given in units of image pixels for both $X$ and $Y$ positions and for an error radius in units of arcseconds: 
$\Delta \mathrm{(RA,Dec)} = \mathrm{pixel size}\, \times\sqrt{\Delta X^2+\Delta Y^2}$.
For derived quantities such as the sum of count rates in multi-band fits or the
hardness ratios of energy bands, the count rate errors are propagated by treating them
like Gaussian errors. 

For each source passing the likelihood thresholds, the following source parameters are written to the output source list:

\begin{itemize}
    \item best-fit position with errors in image, equatorial, and Galactic coordinates;
    \item best-fit source extent value with errors;
    \item best-fit counts, count rates, and fluxes with errors;
    \item likelihood values for detection and extent;
    \item exposure and background map values at the source position;
    \item hardness ratios of the form $HR_{12}= (cr_2-cr_1)/(cr_1+cr_2)$ with errors, \GL{where $cr_1$ and $cr_2$ are the count rates in two of the input energy bands}.
    \item the radius of the \FINAL{sub-region} used for fitting, the PSF fraction covered by that region, and the fraction of pixels with valid detection mask values.
\end{itemize}

The source list contains several rows per source: one per input image, plus summary rows over the energy bands and, \GL{for the optional case where the images of different telescope modules are supplied separately}, a summary over different instruments.

\paragraph{apetool.}

The task \texttt{apetool} has three main functionalities: {\it i)} it produces maps of the size (radius in pixels) of the PSF across the \erosita field of view (PSF maps), {\it ii)} it performs aperture photometry at a set of user-defined positions, and {\it iii)} it generates sensitivity maps following the method described by \cite{Georgakakis08}. 

The PSF of \erosita is modelled using the shapelet basis functions \citep{Refregier03, Refregier_Bacon03}. The shapelet coefficients that describe the distribution of photons at a certain energy and at a given position in the \erosita field of view are stored in a calibration file. To construct the PSF map, the \texttt{apetool} first extracts the GTI from the attitude file. For each time interval, the shapelet coefficients at the positions of a given pixel are retrieved from the calibration file. The average of each shapelet coefficient across all time intervals is used to reconstruct a model of the PSF at the position of interest. This is then used to measure the size of the PSF in pixels at encircled energy fractions (EEF) between 40-95\% in steps of 5\%. The \texttt{apetool} estimates the PSF size in a regular grid of positions in the \erosita field of view. The density of the grid is a trade-off between speed, size of the final PSF map, and an adequate description of the variations in PSF size across the field of view. The default setup is a grid of $21\times21$ positions  along the X and Y direction.

In the case of aperture photometry, there are two options. Counts are either extracted at source positions defined in a source catalogue generated by the \texttt{ermldet} task or at arbitrary user-defined positions. The difference between the two options is the format of the input and output source lists. In the first case, the \texttt{ermldet} source catalogue is supplemented with aperture photometry products. These include {\it i)} the total counts at a given position (source and background) extracted within an aperture of size defined in units of EEF, {\it ii)} the background counts extracted from the source maps generated by the \texttt{ermldet}, {\it iii)} the mean exposure time, {\it iv)} the EEF used to define the extraction radius, {\it v)} the size of the extraction radius in pixels, and {\it vi)} the Poisson probability that the extracted counts (source + background) are a fluctuation of the background. These quantities can be combined to estimate the source flux. The case of aperture photometry at arbitrary source positions corresponds to science applications related to, for instance, X-ray stacking analysis or searches for faint X-ray emission (i.e. below the formal X-ray detection threshold) associated with counterparts selected at other wavelengths or external (non-\erosita) X-ray catalogues. The apertures within which counts are extracted at user-defined positions are expressed in terms of EEF. The quantities as listed above are estimated and stored in an output file.  

The sensitivity map generated by \texttt{apetool} has units of counts. They represent the minimum number of photons (source and background) within the extraction aperture (expressed in EEF units) at a given position on the detector, so that the Poisson false-detection probability is lower than a user-defined value. The Poisson false-detection probability is defined as the probability that the background fluctuats in a random fashion to produce counts within an aperture above a given value. This probability is given by the survival function of the Poisson probability distribution. The resulting sensitivity maps can be combined with the aperture photometry \TL{determined} by \texttt{apetool} for the \texttt{ermldet} sources to provide an accurate representation of the point-source selection function (probability of detecting a source) at a given Poisson false-detection threshold  \citep[see][]{Georgakakis08}. The combination of the \texttt{apetool} sensitivity maps with the aperture photometry of the \texttt{ermldet}-detected sources (also via \texttt{apetool}) enables, among others, the accurate estimation of the X-ray point-source number count distribution as a function of X-ray flux, that is, $\log N-\log S$ (see Section \ref{sec:lognlogs}). 

The sensitivity map FITS file includes three image extensions: {\it i)} the sensitivity map itself, that is, the minimum number of counts as described above as a function of detector position, {\it ii)} the expected background within the extraction radius, and {\it iii)} the corresponding average exposure time (mean of the exposure map) within the extraction radius. With these three components, it is possible to estimate the detection probability of a source with a given count rate or flux. The resulting area or sensitivity curve is stored in the fourth (table) extension of the sensitivity map fits files. It includes count rates and the corresponding area in square degrees within which a source of this count rate can be detected. This calculation makes no assumptions on the X-ray spectral shape of the source.

\paragraph{catprep.}

The task \texttt{catprep} re-formats the  source lists written by the tasks \texttt{ermldet} and
\texttt{apetool}. The output format contains only one row per source; the energy-band specific
values are written as separate columns.
The task also assigns a source identification string containing the root name of the input file,
the source number from the input list, and the processing version.
\section{eROSITA calibration database}
\label{appendix_caldb}

The \texttt{eSASS} calibration database (CALDB) contains all information required for calibrating the data collected with the \erosita telescope modules (TMs). It is based on the HEASARC\footnote{https://heasarc.gsfc.nasa.gov/} CALDB framework for managing calibration files. Each of the seven telescope modules of \erosita has its own set of calibration files. The names of the calibration files follow a naming convention that specifies the telescope module, calibration type, and, if necessary, the type of detector or filter for which this calibration file is intended, as well as the date from which this information is valid and the version number of the file. For example, \texttt{tm2\_badpix\_190712v01.fits} contains the information on the bad pixels of TM2 valid from 12 July 2019 (i.e. before launch) in the first version.

\subsection{Telescope and point spread function}

\paragraph{Vignetting.}
Vignetting is defined here as the flux of a point source contained in a circle of 4\arcmin\ radius relative to that flux at the on--axis position. The vignetting database contains parameters for computing this value for any given photon energy and off--axis angle. 

\KD{These parameters were derived from the same data as were taken to determine the point spread function (see below). Variations in the flux of the X--ray source were considered by performing simultaneous measurements with a monitor counter. This method did not only save measurement time, but also had the advantage that the vignetting centre (i.e. the point of minimum vignetting, which is not necessarily coincident with the point of minimum HEW and which may move with energy) could be determined from the narrow grids of PSF measurements, and that the assumption of azimuthally symmetric vignetting could be verified.}
Comparisons between the measured values and the modelled vignetting curves are shown in Fig.\,7 of \citet{Dennerl2020}.

\paragraph{Point spread function: images.}
In the ground calibration, images of the PSF were obtained at the MPE PANTER facility\footnote{https://www.mpe.mpg.de/heg/panter} individually for each mirror assembly (MA) at the energies of the C--K, Cu--L, Al--K, Ag--L, Ti--K, Fe--K, and Cu--K emission lines, covering the energy range from 0.3 to 8.0~keV. The MAs were placed 124~m away from an X--ray point source and mounted together with the X--ray camera TRoPIC on a rigid platform, which could be tilted horizontally and vertically. The TRoPIC X--ray CCD is identical in its pixel size to the eROSITA CCDs, but consists of only $256\times256\mbox{ pixels}$. In order to sample the full eROSITA focal plane, TRoPIC was mounted on a manipulator, which shifted it to five different positions to cover the central and the four outer parts of the \FINAL{field of view}. The four outer TRoPIC positions sampled the PSF on a rectangular $6'\times6'$ grid in 121 images, covering the offset angles from $-30'$ to $+30'$. The central TRoPIC position measured the PSF in 36 images on a $6'\times6'$ grid that was shifted by $3'$ with respect to the outer grid and covered the off-axis angles between $-15'$ and $+15'$. This resulted in 157 images per MA and energy. In order to save measurement time, Cu--L exposures were only done for TM1 and TM2, and the outer part of the \FINAL{field of view} was only sampled in the upper right quadrant at Ag--L, Ti--K, Fe--K, and Cu--K, except for TM1, where the full grid was covered in all the seven energies.

In order to obtain the required large number of PSFs within a reasonable time, the measurements were taken at very high photon rates, taking extreme pile--up into account and making use of the fact that, if the dominant incident energy is known, the number of photons per pixel can be reconstructed from the total charged released there. In this mode, however, it is not possible to apply the usual technique to identify and suppress the traces of minimum ionising particles by selecting only valid pixel patterns, because photon pile--up may also create `invalid' pixel patterns. This problem could be solved by developing specific criteria for spotting suspicious pixel patterns, by removing all the contaminated CCD frames, and by visually checking the remaining frames for each of the 4860 PSF exposures. It turned out that only $\sim5\%$ of the frames needed to be rejected. In this way, the exposure time per PSF could be reduced from several hours to $\sim80$~seconds, boosting the efficiency by two orders of magnitude.

The further processing of the PSFs required transforming their location on TRoPIC to that on the eROSITA focal plane. This made it necessary to calibrate the geometrical properties of the PANTER setup, which consists of several distances and angles. In total,
seven geometrical quantities needed to be determined.

\begin{figure}
    \centering
    \includegraphics[width=0.45\textwidth]{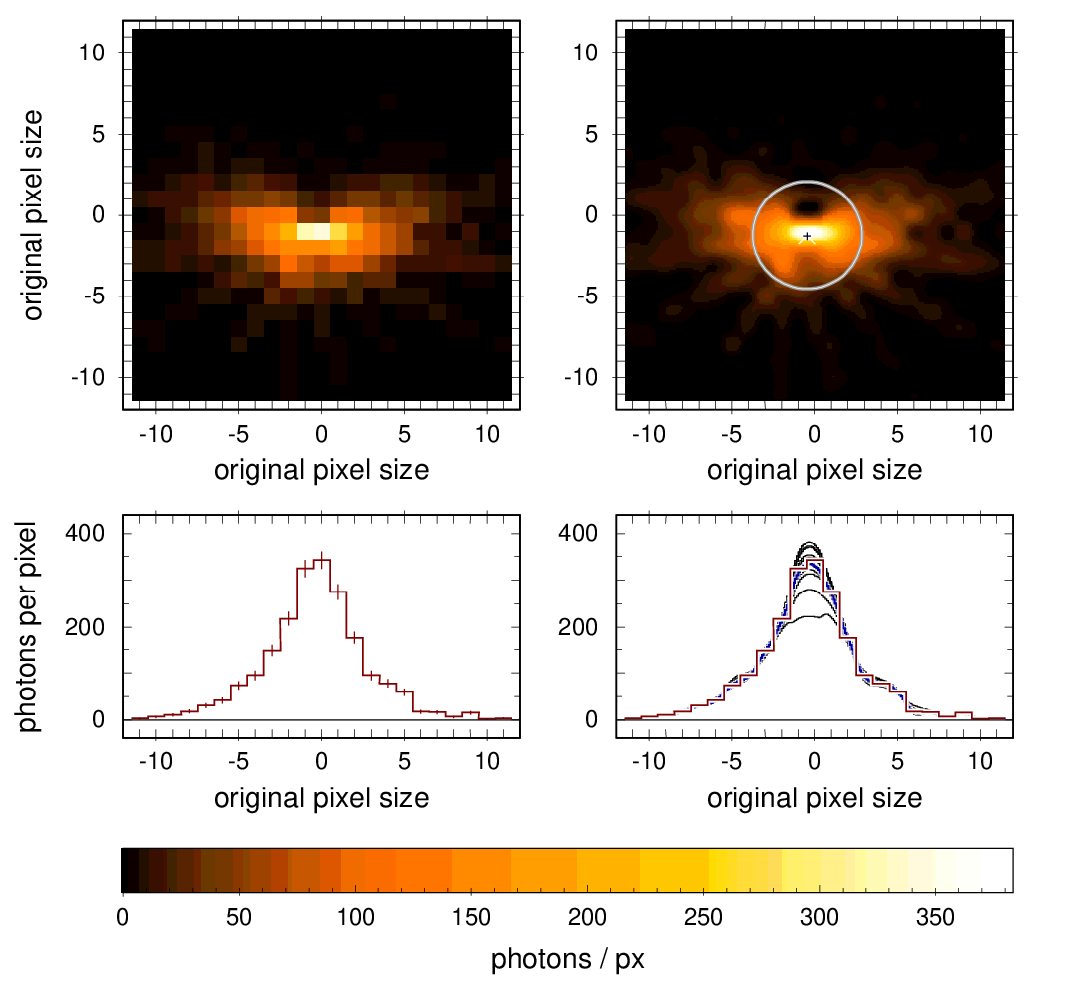}
    \caption{Example of the 1.5~keV TM1 PSF at an off--axis angle of $30'$ in the original resolution (left) and the reconstructed resolution (right), where each original pixel is \FINAL{sub-divided} into $10\times10$ \FINAL{sub-pixels}. The histograms at the bottom show the flux distribution in the brightest pixel row (left) and in the ten corresponding \FINAL{sub-pixel} rows (right). The modified bicubic resampling method enhances fine structures. The white circle indicates the minimum HEW.
    }
    \label{fig:psf_rebin}
\end{figure}

Another challenge was that the PSFs were available only at the native $9\farcs5$ resolution provided by the TRoPIC pixel size (\FINAL{sub-pixel} resolution could not be used here because of the high pile--up). In order to increase the spatial resolution, we tried to find plausible flux distributions that agreed with the measured ones after binning them into $9\farcs5 \times 9\farcs5$ pixels. This was done by developing a modified method of bicubic spline interpolation, which was made to be flux conservative (Fig.\,\ref{fig:psf_rebin}). These images are the basis for modelling the PSF. More details are presented in \cite{Dennerl2020}.

\paragraph{Point spread function: shapelet model.}

\begin{figure}
    \centering
    \includegraphics[width=\columnwidth]{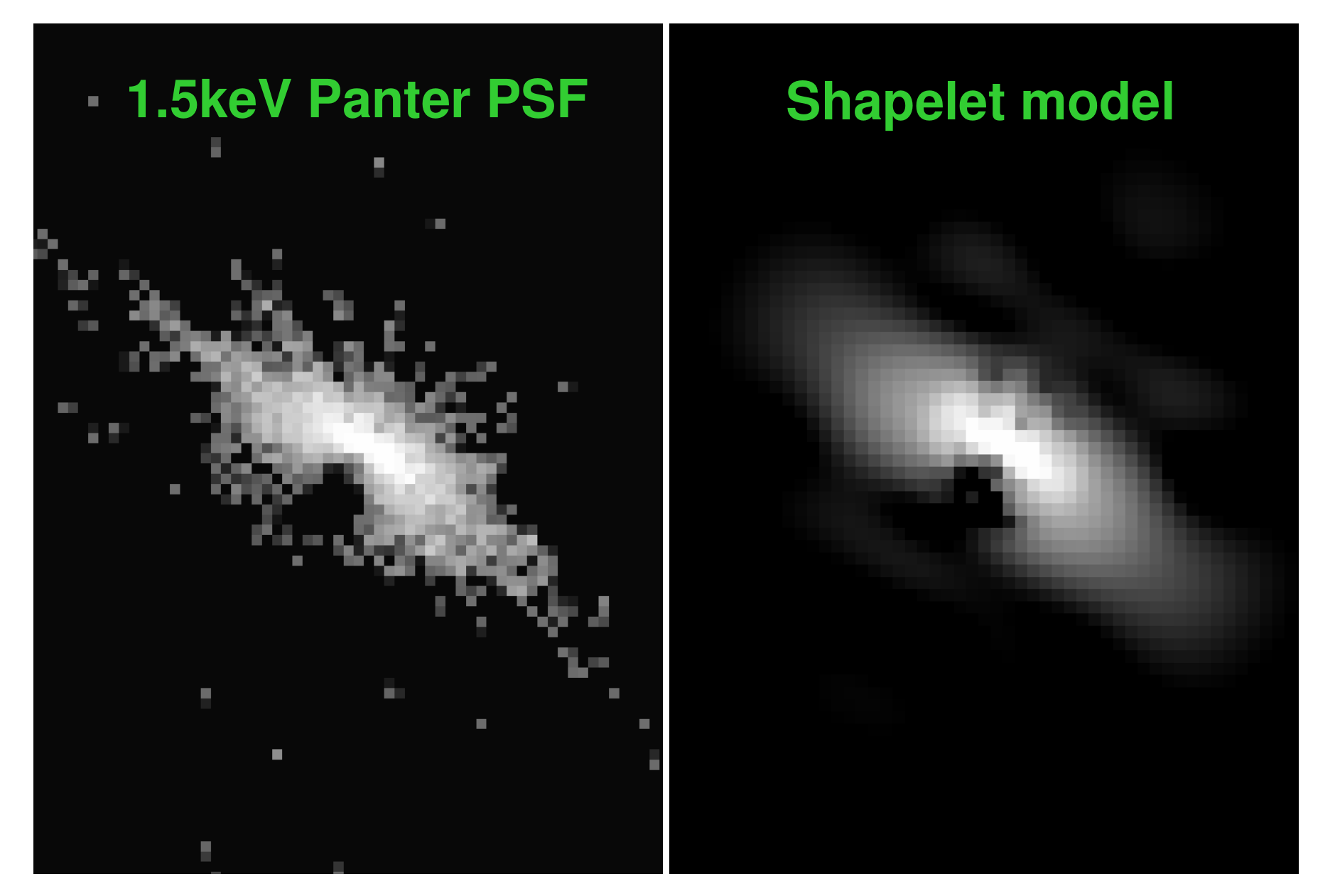}
    \caption{Example of the 1.5\,keV TM1 PSF at an off-axis angle of about \LAST{26}\arcsec. The image on the left shows the PANTER calibration image. On the right, we show the PSF reconstruction using the shaplet model described in the text.}
    \label{fig:shapelib}
\end{figure}

The \erosita optics and large field of view translate into a complex PSF shape that depends on both off-axis angle and energy. The regular nearly Gaussian PSF on-axis is distorted with increasing off-axis angle, leading to elongated features and asymmetries (see Figure  \ref{fig:shapelib}) that are hard to model using analytic functions. This has motivated an approach whereby the PSF at a given position is linearly decomposed into an appropriately chosen two-dimensional set of localised basis functions. The PSF can then be represented by the coefficients of the basis function components. The decomposition uses the shapelet functions proposed by \cite{Refregier03, Refregier_Bacon03}, which represent  a complete and orthonormal set in the two-dimensional space. The shapelets are weighted Hermite polynomials and correspond to perturbations of a circular Gaussian profile. In one dimension, these are described by the relation

\begin{equation}
    \phi_\nu (x; \beta) = \left[ 2^n\,\beta \pi^\frac{1}{2}\,n! \right] ^{-\frac{1}{2}} \, H_{n}(x/\beta)\,e^{-\frac{x^2}{2\,\beta^2}},
\end{equation}

\noindent where $n$ is a non-negative integer, $H_n(x)$ is the Hermite polynomial of order $n,$ and $\beta$ is the scale of the shapelet function. In the case of two-dimensional images with coordinates $(x,y),$ the shapelet basis functions are given by the relation

\begin{equation}
    \psi_{\nu, \mu}(x, y; \beta) =  \phi_\nu (x; \beta) \cdot  \phi_\mu (y; \beta).
\end{equation}

\noindent In this case, the order of the shapelet function is characterised by the two non-negative integers $n$, $m,$ and there is a single scale $\beta$ for both  $\phi_\nu (x)$, $\phi_\mu (y)$. The choice of the scale $\beta$ depends on the size of the features to be modelled. The motivation for using this set of functions is that they have been used to reconstruct the complex shapes of galaxies in optical surveys for weak-lensing applications \citep{Massey07}. 

In the case of the \erosita PSF, we used three independent sets of shapelet functions, $\psi_{\nu, \mu}(x, y; \beta)$, each of which corresponds to a different scale. They are intended to model the core, the main body, and the wings of the PSF. The scales of each of the three components were fixed to $\beta=1$, 1.5, and 6\,pixels of the ground-calibration images, which have a pixel scale of $9\farcs5$. The number of shapelet orders is set by the requirement $n+m \le N_\mathrm{max}$, where $N_\mathrm{max}=0$, 10, and 8 for each of the three scales $\beta=1$, 1.5, and 6\,pixels, respectively. The choice of $N_\mathrm{max}=0$ for the smallest scale translates into a single Gaussian function to model the core of the PSF. There are 66 and 45 shapelet coefficients for  $N_\mathrm{max}=10$ (scale $\beta=1.5$\,pixels) and $N_\mathrm{max}=8$ (scale $\beta=6$\,pixels). The total number of coefficients (free parameters) that describe the PSF shapelet model at a given position on the detector and energy is 112. The choice of the $\beta$, $N_\mathrm{max}$ values is the result of experimentation and represents the minimum number of scales and coefficients that provide a reasonable representation of the PSF at all off-axis angles and energies (see  Figure  \ref{fig:shapelib}).  Fixing the scale and $N_\mathrm{max}$ of the shapelets enables the \FINAL{co-addition (stacking)} of PSFs across energies and off-axis angles in the shapelet-coefficient space rather than the image-pixel space and significantly accelerates (factors of 10-100) the calculation of the model PSFs.

\AGE{The PANTER PSF images for different energies and off-axis angles described earlier in this section with a pixel size of $9\farcs5$ were fit with the shapelet model above to determine the maximum likelihood coefficients (112 for each PSF image). The modelled energies were 0.3, 1.5, 2.0, 4.5, 5.4, and 6.4\,keV.}
Figure  \ref{fig:shapelib} shows an example PSF image taken at PANTER for the TM1 module for photons with energy 1.5\,keV. For comparison, the shapelet-model reconstruction of the PSF is also presented in this figure. \AGE{The fraction of photons predicted by the shapelet model within a fixed radius is typically within 10-20\% of the PANTER measurement. This difference is related to the maximum adopted shapelet scale of 6\,pixels ($\approx1$\,arcmin). Photons beyond this scale cannot be reproduce by the current model.} The shaplet coefficients were packed in calibration files that were accessed by the various \texttt{eSASS} tasks. For a given input position on the \erosita detectors and a given energy, the shapelet coefficients were estimated using linear interpolation in three dimensions (two space dimensions and one energy dimension).    

\subsection{Pattern recombination and energy calibration}

While no calibration data are needed for the pattern recombination, this is very different for the energy calibration because each CCD column is characterised by an individual CTI and gain, and these values depend on the energy, the CCD temperature, and the time of the observation. The initial database, which did not yet consider the last two dependences, contains already 21616 parameters.
\KD{In the updated currently used database, which considers the effects of the CCD temperature and the time of the observation, the number of parameters has increased to 29764.}
Another database, consisting of a total of 1799 \FINAL{sub-pixel} maps, is used for a fast reconstruction of the \FINAL{sub-pixel} position from the distribution of the charge over the pattern \citep{Dennerl2012}.

\subsection{Detector and camera characterisation}

\paragraph{Bad pixels.}
The bad-pixel calibration files characterise individual bad pixels and groups of bad pixels (specified as rectangles) of each camera by type, affected amplitude range, and time at which they were active. The list of bad pixel is updated continuously as the behaviour of the detectors evolves. 

\paragraph{Bad times.}
These calibration files contain the start and end times of periods when data are either unavailable or cannot be used scientifically because of a malfunction. In addition, time periods are listed in which a TM is offline due to a camera reset \citep{Predehl2021} or is switched of, for instance, because of an orbit correction. In the calibration and performance verification (Cal-PV) phase, these time periods were set for TM3 and TM6.

\paragraph{Time shifts.}
Both the Interface and Thermal Controller (ITC) and the Camera Electronics (CE) store a local copy of the on-board time source (OTS) counter \citep{Predehl2021}. These local copies are incremented by a 1~Hz pulse signal from the spacecraft.  ITC and CEs can only be updated with the OTS time word from the spacecraft when they are in a specific mode. For the time being, this is only the case when a CE is reset or switched on. On these occasions, it can happen that the time counter of the CE jumps by 1\,s. 

\SF{Time shifts can normally only be detected in survey or field-scan mode (if the scan velocity is sufficiently high) because data from CEs with incorrect time synchronisation are then projected at incorrect sky locations. With pointed observations, such as those made predominantly in the Cal-PV phase, each event is projected onto the same position in the sky, and time shifts can only be detected by observing a stable clock such as a pulsar. It is known from pulsar observations during the Cal-PV phase that a few time shifts occurred. The only time shift during the Cal-PV phase that could be clearly assigned is from November 24, 2019, for TM2. The \texttt{timeoff} calibration files contain the date and the direction of each time shift ($0\rightarrow 1$ or $1\rightarrow 0$) for each camera from this date on. Time shifts applicable for each camera before this date were determined directly from the eFEDS data and from two all-sky survey test scans performed on September 23 and November 15, 2019. The TM6 time shift in eFEDS sub-field I reported in \S~2 of the main paper is not corrected.}

\paragraph{Detector maps.}
The detector maps are intended to hold fractional sensitivity values of each detector pixel. While the detector maps are evaluated by eSASS tasks \texttt{expmap} and \texttt{srctool}, it is currently not foreseen to make use of this calibration mechanism, and all pixel values are set to 1.0.

\paragraph{FOV maps.}
The \FINAL{field of view} maps define the field of view of each camera. Pixels outside the mask will receive the \texttt{OUT\_OF\_FOV} flag (set by task \texttt{evprep}), they will not be projected onto the sky, and the corresponding areas will be excluded from the exposure maps. 

\subsection{Boresight}

\paragraph{Star trackers and gyroscopes.}

For each of the four attitude sources of the SRG satellite (SED26/1, SED26/2, BOKZ, and Q-Gyro) a calibration file is maintained. Each file holds an orientation quaternion defining the transformation between the coordinate system of the respective device and the nominal camera orientation of telescope module TM1. 

\paragraph{X-ray telescopes.}

The geometrical parameters for each telescope module are stored in the respective instrument calibration files.
Each file contains the plate scale and the camera boresight with respect to the nominal orientation of TM1, the boresight is stored in the form of Euler angles (pointing offsets in X, Y, and roll angle).
\section{eROSITA standard calibrated data products}
\label{appendix_products}
The \erosita data analysis pipeline provides a set of calibrated data products described below. All products are FITS files complying (where feasible) with established standards such that in addition to eSASS, a range of general-purpose astronomical data analysis tools may be used. A full description of the main product files is available online\textsuperscript{\ref{webdoc}}.  

\subsection{Calibrated event files}
\erosita calibrated event files are containers that provide a full set of X-ray event and auxiliary data required for most data analysis tasks in multiple FITS extensions. They consist of one \texttt{EVENTS} FITS table extension holding data from one or several \erosita cameras as well as various camera specific extensions.  

\paragraph{X-ray events.}
The \texttt{EVENTS} FITS table extension holds information on each observed event and for recombined, calibrated X-ray photons. Event coordinates are provided in units of detector and sky pixel coordinates and as right ascension and declination. Further information includes event arrival times, raw and calibrated event amplitudes, a variety of event pattern type information (see Appendix \ref{appendix_esass}.1, {\tt pattern}), and a set of event flags characterising each event.

\paragraph{Good time intervals.}
Good-time intervals (GTI) are provided for each camera in the \texttt{GTIn} FITS table extensions, where n is the TM number. The GTIs indicate the time periods when each camera was operational and collecting event data. Optionally, a set of \texttt{FLAREGTIn} FITS table extensions is included. They specify low-background time periods free of background flares (see Appendix \ref{app:photometry}, \texttt{flaregti}).  

\paragraph{Live time.}
The \texttt{DEADCORn} FITS table extensions provide the fractional live time for each 50~ms time interval of each camera. The \texttt{DEADORn} extension tracks the fraction of the detector area that is not available for the detection X-ray events due to minimum ionising particles (MIPs) hitting the detector. Where applicable, the \texttt{DEADCORRn} extension also records exposure losses when a fraction of the exposure time is intentionally discarded, for instance to prevent count rates exceeding technical limitations.    

\paragraph{Telescope specific attitude.}
Right ascension, declination, and roll angles with respect to the camera \texttt{(RAWX, RAWY)} coordinate system are provided for each camera in the \texttt{CORRATTn} FITS table extensions with a time resolution of one second. 

\paragraph{Bad pixel information.}
The \texttt{BADPIXn} FITS table extensions list and characterise bad pixels of each camera. The information provided includes the affected energies and time intervals as well as the bad-pixel type (bright, masked on-board, etc.). The format of the bad-pixel event file extensions is closely related to the bad-pixel calibration files described in Appendix \ref{appendix_caldb}.3.

\paragraph{Housekeeping information.}
For each camera, four FITS table extensions are provided that hold a subset of the \erosita housekeeping data required for event calibration and exposure calculation, such as the observing mode and filter wheel position of each camera.

\subsection{Count images and maps}

\paragraph{Images.}
The \texttt{eSASS} science images are created by the task \texttt{evtool} and are stored in the primary extension of a FITS file. The FITS header contains the standard \texttt{eSASS} keywords copied from the original event file, keywords describing the event selection, and a World Coordinate System (WCS) header.
By default, the FITS file also contains the EVENTS and auxiliary extensions from the original events file with the same selections applied as were used to create the image. 

\paragraph{Exposure maps.}

The exposure maps contain the results of the exposure calculation performed by the task \texttt{expmap}.
The exposure maps can either contain the raw  exposure times for each map pixel (unvignetted exposure)
or the exposure multiplied with the energy and off-axis angle-dependent telescope vignetting function (vignetted exposure). In both cases, the unit of the pixel values is seconds. 
The WCS header is identical with that of the input template image.

\paragraph{Background maps.}

The background maps contain the source free sky $+$ detector background as determined
by the task \texttt{erbackmap}. The values are stored in units of counts/pixel, and the 
image size and WCS coordinate system of a background map is identical with that of the respective science image.

\paragraph{Sensitivity maps.}

The sensitivity maps contain limiting fluxes for sources that can be detected in an \erosita observation.
The nominal unit of the pixel values is $\mathrm{ erg/(s\; cm^2)}$, but it depends on the 
energy conversion factor provided for the task.

\subsection{Source level products}

\paragraph{Source catalogues.}

Source catalogues are written as FITS tables by the tasks \texttt{ermldet} and  \texttt{apetool} 
and are then re-formatted by the task \texttt{catprep}. The final format contains one row per source, with a column for each of the source parameters described in section  \ref{app:photometry}.

\paragraph{Spectra.}
Spectra are produced by the task \texttt{srctool} and are standard OGIP-compliant\footnote{OGIP FITS Working Group, https://heasarc.gsfc.nasa.gov/} data products.
The task writes output spectra for each TM and a combined spectrum.
Each spectrum consists of the number of counts in an extraction region within each PI channel after filtering has been applied.
There are some subtleties in how \texttt{srctool} defines various output quantities due to the scanning nature of the \erosita telescope.
The \texttt{ONTIME} is the total amount of time for which the source is within the field of view during the input GTIs.
The time intervals within the \texttt{ONTIME} (source specific GTIs) are normally written to the \texttt{srctool} output products as GTI extensions.
The \texttt{EXPOSURE} is the \texttt{ONTIME} reduced to account for dead time fraction given in the input event file.
The \texttt{BACKSCAL} is the average area on the sky of the source or background extraction region during the source-specific GTIs.
For background subtraction, it is often useful to know the geometric area on the sky of the extraction region, which is written as a \texttt{srctool}-specific keyword called \texttt{REGAREA}.
The geometric area of where the source model is non-zero is written as \texttt{RGDMAREA}.
In addition to the standard \texttt{COUNTS} column in the output spectra, srctool also writes columns containing the number of single, double, triple or quad patterns in each channel.
The total number of counts of each type is written into keywords as \texttt{CNTS\_S}, \texttt{CNTS\_D}, \texttt{CNTS\_T,} and \texttt{CNTS\_Q}, and the total as \texttt{CTS}.
When combining spectra, srctool adds the counts and computes average \texttt{ONTIME} and \texttt{EXPOSURE} values, while  \texttt{BACKSCAL},  \texttt{REGAREA,} and \texttt{RGDMAREA} are exposure-weighted averages.

\paragraph{Light curves.}
The output light-curve files contain the binned X-ray light curve in several bands, following OGIP recommendations for time-analysis data files.
The columns include the midpoint of the time bin (\texttt{TIME}), the width of the bin (\texttt{TIMEDEL}; $\delta_T$), the number of counts in each band (\texttt{COUNTS}; $c$), the number of counts in the background region in each energy band (\texttt{BACK\_COUNTS}; $c_{B}$), the fraction of the nominal on-axis effective collecting area computed in each band (\texttt{FRACAREA}; $f_A$), the fraction of the time bin that overlaps with input GTIs and when the source was visible in the field of view (\texttt{FRACTIME}; $f_T$), the product of the fractional collecting area and fractional temporal coverage (\texttt{FRACEXP}; $f_E = f_A f_T$), the mean off-axis angle during the bin, and the ratio by which the background counts need to be scaled to give the background contribution in the source \TL{aperture} (\texttt{BACKRATIO}, $r$).
The \texttt{RATE} column is calculated from the other columns in each band as $(c-r\, c_{B})/(f_E \, \delta_T)$.
The $1\sigma$ uncertainty on the rate is given for each energy band in \texttt{RATE\_ERR}, providing there are more than 25 counts in the respective counts column, using $\sqrt{c + r \, c_{B}}/(f_E \, \delta_T)$.

\subsection{Detector response and effective areas}

\paragraph{Response matrices.}
Response matrices (RMFs) generated by \texttt{srctool} are OGIP-format response files, containing standard \texttt{MATRIX} and \texttt{EBOUNDS} extensions.
The output depends on the pattern selection chosen by the user.
Output response matrices are the sum of the response matrices for the different patterns.
Similarly to spectra, \texttt{srctool} writes response matrices for the individual TMs and for the combination.
The combined response matrix is the exposure-weighted average of the individual response matrices.

\KD{The RMFs were derived from measurements that had been performed with a prototype version of the eROSITA CCDs at the synchrotron radiation facility BESSY II, where the energy range 0.08\,--\,11.0~keV was sampled with monochromatic measurements at 29 energies \citep{Granato2010}. The measurements were cleaned from artefacts (e.g. higher orders), filtered to contain only single-pixel events, and then used to derive an empirical mathematical function that reproduced the observed spectra. After this function was available, it was straightforward to evaluate it at any energy and to compose an empirical RMF. This method was also applied to the other three valid pattern types (doubles, triples, and quadruples) by using the same generic function and modifying only some parameters. This resulted in four initial RMFs, one for each pattern size \citep{Dennerl2020}. These are the RMFs that are currently used for all TMs. It is planned to adapt these generic RMFs to the properties of the individual TMs by using dedicated calibration observations of suitable astrophysical targets. In this way, it should also be possible to take temporal variations into account.}

\paragraph{Auxiliary response files.}
The auxiliary response files (ARFs) made by \texttt{srctool} are OGIP-compliant. 
These files include several different effects that are applied to the on-axis effective area, including vignetting, PSF correction, and corrections for times at which the source is only partially visible.
The ARF files contain a \texttt{SPECRESP} extension that in turn contains the standard columns of \texttt{ENERGY\_LO}, \texttt{ENERGY\_HI} , and  \texttt{SPECRESP} , which give the effective area curve.
In addition, \texttt{srctool} writes columns for diagnostic purposes that give the corrections applied to the on-axis effective area due to PSF (as \texttt{CORRPSF}), vignetting (as \texttt{CORRVIGN}), and the two combined (\texttt{CORRCOMB}).
\section{Description of catalogue entries}
\label{appendix_catalog}

\begin{table}[htp]
  \centering 
  \caption{Energy bands}
  \begin{tabular}{lll}
    \hline
    {\sl Band} & energy range  & ECF \\
               & keV & cm$^2$/erg \\
    \hline
    \multicolumn{3}{l}{single-band detection}\\
    \hline
              &  0.2--2.3      & 1.074$\times 10^{12}$\\
    \hline
    \multicolumn{3}{l}{three-band detection}\\
    \hline
 1            &  0.2--0.6      & 1.028$\times 10^{12}$\\
 2            &  0.6--2.3      & 1.087$\times 10^{12}$\\
 3            &  2.3--5        & 1.147$\times 10^{11}$\\
    \hline
    \multicolumn{3}{l}{post-hoc photometry}\\
    \hline
 s            & 0.5--2         & 1.185$\times 10^{12}$\\
 \LAST h            & 2.3--5         & 1.147$\times 10^{11}$\\
 u            & 5--8           & 2.776$\times 10^{10}$ \\
 b1           & 0.2--0.5       & 9.217$\times 10^{11}$\\
 b2           & 0.5--1         & 1.359$\times 10^{12}$\\
 b3           & 1--2           & 1.014$\times 10^{12}$\\
 b4           & 2--4.5         & 1.742$\times 10^{11}$\\
    \hline
\end{tabular}
 \tablefoot{ECF: Energy conversion factors for the four source-detection bands and the seven forced-photometry bands.}
\label{table:bands}
\end{table}

\begin{table*}[htp]
  \centering
   \caption{Catalogue column description}
  \begin{tabular}{p{0.21\textwidth} p{0.04\textwidth} p{0.08\textwidth} p{0.56\textwidth}}
   \hline
Column         &Format &Units & Description \\
   \hline
    \multicolumn{4}{l}{1. Source properties from PSF-fitting detection}\\
   \hline
Name           &22A     &..     & \\
ID\_SRC         &J      &..     & Source ID\\
\FINAL{ID\_main (ID\_hard)}  &J  &.. & \FINAL{Counterpart source ID in the main (hard) catalogue for sources in the hard (main) catalogue}\\
RA             &D       &deg    & Right ascension (ICRS), uncorrected\\
DEC            &D       &deg    & Declination (ICRS), uncorrected \\
RADEC\_ERR      &E      &arcsec & Combined positional error, uncorrected \\
RA\_CORR        &D      &deg    & Right ascension (ICRS), corrected\\
DEC\_CORR       &D      &deg    & Declination (ICRS), corrected\\
RADEC\_ERR\_CORR &D     &arcsec & Combined positional error, corrected\\
EXT            &E       &arcsec & Source extent parameter\\
EXT\_ERR        &E      &arcsec & Extent error\\
EXT\_LIKE       &E      &..     & Extent likelihood\\
DET\_LIKE$_\ast$       &E       &..     & Detection likelihood\\
ML\_RATE$_\ast$        &E       &cts/s  & Source count rate measured by PSF-fitting\\
ML\_RATE\_ERR$_\ast$    &E      &cts/s  & 1-$\sigma$ count rate error\\
ML\_CTS$_\ast$         &E       &cts    & Source net counts measured from count rate\\
ML\_CTS\_ERR$_\ast$     &E      &cts    & 1-$\sigma$ source counts error\\
ML\_FLUX$_\ast$        &E       &erg/cm$^2$/s   & Source flux in the detection band \\
ML\_FLUX\_ERR$_\ast$    &E      &erg/cm$^2$/s   & 1-$\sigma$ source flux error\\
ML\_EXP$_\ast$         &E       &s      & Vignetted exposure time at the source position\\
ML\_BKG$_\ast$         &E       &cts/arcmin$^2$ & Background at the source position\\
inArea90        &L  &  &True if inside the inner region with 0.2-2.3 keV vignetted exposure above 500s, which comprises 90\% of the total area\\
\hline
    \multicolumn{4}{l}{2. Forced PSF-fitting results for seven energy {\sl Band}s (Table.~\ref{table:bands}.1); 7$\times$15 columns} \\
    \hline
DET\_LIKE\_{\sl\small Band}   &D        &..     & Detection likelihood \\
ML\_RATE\_{\sl\small Band}    &D        &cts/s  & Source count rate \\
ML\_RATE\_ERR\_{\sl\small Band}&D       &cts/s  & 1-$\sigma$ combined count rate error \\
ML\_RATE\_LOWERR\_{\sl\small Band}&D    &cts/s  & 1-$\sigma$ lower count rate error \\
ML\_RATE\_UPERR\_{\sl\small Band}&D     &cts/s  & 1-$\sigma$ upper count rate error \\
ML\_CTS\_{\sl\small Band}     &D        &cts    & Source net counts \\
ML\_CTS\_ERR\_{\sl\small Band} &D       &cts    & 1-$\sigma$ combined counts error \\
ML\_CTS\_LOWERR\_{\sl\small Band} &D    &cts    & 1-$\sigma$ lower counts error \\
ML\_CTS\_UPERR\_{\sl\small Band} &D     &cts    & 1-$\sigma$ upper counts error \\
ML\_FLUX\_{\sl\small Band}    &D        &erg/cm$^2$/s   & Source flux \\
ML\_FLUX\_ERR\_{\sl\small Band}&D       &erg/cm$^2$/s   & 1-$\sigma$ combined flux error \\
ML\_FLUX\_LOWERR\_{\sl\small Band}&D    &erg/cm$^2$/s   & 1-$\sigma$ lower flux error \\
ML\_FLUX\_UPERR\_{\sl\small Band}&D     &erg/cm$^2$/s   & 1-$\sigma$ upper flux error \\
ML\_EXP\_{\sl\small Band}     &D        &s      & Vignetted exposure time at the source position\\
ML\_BKG\_{\sl\small Band}     &D        &cts/arcmin$^2$ & Background at the source position \\
\hline
    \multicolumn{4}{l}{3. Aperture photometry results for seven energy {\sl Band}s (Table.~\ref{table:bands}.1); 7$\times$5 columns}\\
    \hline
APE\_CTS\_{\sl\small Band}    &J        &cts    & Total counts extracted within the aperture\\
APE\_EXP\_{\sl\small Band}    &D        &s      & Exposure map value at the given position\\
APE\_BKG\_{\sl\small Band}    &D        &cts    & Background counts extracted within the aperture, excluding nearby sources using the source map\\
APE\_RADIUS\_{\sl\small Band} &D        &pixels & Extraction radius\\
APE\_POIS\_{\sl\small Band}   &D        &..     & Poisson probability that the extracted counts (APE\_CTS) are a background fluctuation\\

\hline  
  \end{tabular}
  \tablefoot{
    The suffix in the column name of forced PSF fitting and aperture photometry results (section 2 and 3)  indicates the energy band (Table.~\ref{table:bands}). 
    The columns of the three-band detected hard catalogue are almost identical to those of the single-band detected main and supplementary catalogues, except that the PSF-fitting output columns (those marked with $a$ subscript asterisk) have four sets of data with an energy-band suffix (0, 1, 2, and3) in the column name rather than only one set of data without this energy-band suffix (as listed in section 1).
    \FINAL{Moreover, the ID\_main and ID\_hard columns are only for the hard and main catalogues, respectively.}
    }
      \label{table:columns}
\end{table*}

\LAST{The catalogues are available with this paper and on the eROSITA Early Data Release website\footnote{https://erosita.mpe.mpg.de/edr/eROSITAObservations/Catalogues/}.}
Table~\ref{table:bands} lists the energy bands and corresponding ECFs used in this work. The ECFs are calculated assuming an absorbed power law with a slope of 2.0 and with a Galactic absorbing column density of 3$\times 10^{20}$ cm$^{-2}$.

The main and the supplementary catalogues from the 0.2--2.3 keV single-band detection share the same columns as described in Table~\ref{table:columns}.
\TL{The seven-band forced photometry is done independently for the single-band and three-band detected sources.
In the case of the three-band detected hard catalogue, four sets of output columns from the source detection PSF-fitting (not the forced PSF-fitting) have the same names as listed in section 2 of Table~\ref{table:columns} , but with an energy-band suffix in the column name indicating the three source-detection bands (1, 2, and 3) or the summary of the three bands (0).}
The only exception is the ML\_EXP\_{\sl\small Band}, which is only for the three individual bands (1, 2, and 3), but not for the summary case (0).
The hard catalogue has an additional column ID\_main, which stores the source ID in the main catalogue for the overlapping sources.

\TL{
The current versions of the catalogues contain a small number of NULL values in the PSF-fitting error measurement columns, for example, ML\_RATE\_ERR, RADEC\_ERR\_ERR, and EXT\_ERR, because in some cases the \texttt{ermldet} algorithm that determines the parameter errors failed to converge. When the determination of the positional error failed (490 sources in the main catalogue; 1.8\% of the total), a mean \texttt{RADEC\_ERR\_CORR} of $4\farcs0$ was assumed, which is the median value of \texttt{RADEC\_ERR}.}\section{eFEDS simulations}
\label{sec:simulation}
In order to inspect and optimise the source detection for \erosita, we \TL{ran} extensive simulations as described in \citet{Liu2021_sim}. The simulation of eFEDS is briefly summarised here.

We \TL{used} \texttt{sixte-2.6.2}\footnote{https://www.sternwarte.uni-erlangen.de/research/sixte/} to create mock eFEDS event files and \TL{implemented} 18 realisations.
To make them as representative as possible to the real data, we input the real \erosita calibration files and eFEDS attitude file, and AGN, star, and cluster catalogues generated from cosmological simulations \citep{Comparat2020}. We \TL{measured} the mean X-ray and particle background spectra from the eFEDS data and \TL{simulated} the source signal, X-ray background, and particle background separately to account for their different vignetting.
We \TL{ran} the eFEDS source detection pipeline on the mock data and then \TL{associated} the detected sources with the input on the basis of the source-ID flag on each photon. 
By checking the number of photons contributed by each input source in the core region of a detected source, we \TL{identified} its primary and secondary (if existed) input counterparts.
The point sources (AGN and stars) and extended sources (clusters) \TL{were} considered separately.
In this way, we \TL{were able to identify} not only the cases where one input source was uniquely matched to one detected source that \TL{was correctly} classified (as point- or extended source) or not, but also the cases where one input source \TL{was} detected as multiple sources and the cases where one detected source \TL{was} due to a blending of multiple input sources.

Based on the input-output association, we \TL{were able to measure the source detection efficiency, that is, the completeness (detected fraction) and contamination level (fraction of spurious sources). The final source detection strategy was chosen according to the impact on source detection efficiency caused by} any adjustment of the source detection procedure and parameters.

\end{appendix}


\end{document}